\pdfoutput=1

\documentclass[letterpaper, 10 pt, conference]{ieeeconf}

\usepackage{graphicx}
\usepackage{booktabs}
\usepackage{adjustbox}
\usepackage{wrapfig}
\usepackage[labelfont=bf,figurename=Fig.,font=small]{caption}
\usepackage{subcaption}
\usepackage{lipsum}
\usepackage{amssymb}
\usepackage{amsmath}
\usepackage{siunitx}
\usepackage{mathtools}
\usepackage{xcolor}
\usepackage[colorlinks = true,
            linkcolor = blue,
            urlcolor  = blue,
            citecolor = blue,
            anchorcolor = blue]{hyperref}
\let\labelindent\relax
\usepackage[inline]{enumitem}
\usepackage[ruled,vlined,linesnumbered]{algorithm2e}
\usepackage[noend]{algpseudocode}
\usepackage{ifthen}
\usepackage{dsfont}
\usepackage[backend=biber,style=numeric-comp,sorting=none,maxbibnames=99]{biblatex}
\usepackage{balance}
\usepackage{dirtytalk}
\usepackage{afterpage}
\usepackage{times}

\allowdisplaybreaks

\newcommand{\E}{\mathbb{E}}

\newcommand{\N}{\mathbb{N}}

\newcommand{\Prob}{\mathbb{P}}

\newcommand{\ra}{\rightarrow}
\newcommand{\R}{\mathbb{R}}
\newcommand{\Ra}{\Rightarrow}

\newtheorem{remark}{Remark}

\newtheorem{proposition}{Proposition}
\newtheorem{definition}{Definition}
\newtheorem{lemma}{Lemma}
\newtheorem{theorem}{Theorem}

\newcommand{\graphOrig}{G_O}
\newcommand{\graphMod}{G}
\newcommand{\nodesOrig}{I_O}
\newcommand{\nodesMod}{I}
\newcommand{\arcsOrig}{A_O}
\newcommand{\arcsMod}{A}

\newcommand{\arcsLoadConstraintSet}{\mathcal{W}}
\newcommand{\arcLoadMod}{w}
\newcommand{\arcLoadOrig}{w}
\newcommand{\arcLoadModDiscrete}{W}
\newcommand{\latency}{s}

\newcommand{\costToGo}{z}
\newcommand{\nodeCostToGo}{\varphi}
\newcommand{\routeSetMod}{\textbf{R}}
\newcommand{\routes}{\routeSetMod}

\newcommand{\depth}{\ell}
\newcommand{\depthNode}{\bar \ell}
\newcommand{\height}{m}
\newcommand{\heightNode}{\bar m}
\newcommand{\probDist}{\xi}

\newcommand{\nodeLoadIn}{g}

\usepackage{todonotes}
\usepackage{bbm}

\bibliography{references}

\IEEEoverridecommandlockouts
\title{\LARGE \bf Arc-based Traffic Assignment: Equilibrium Characterization and Learning
}

\author{Chih-Yuan Chiu$^{\star 1}$, Chinmay Maheshwari$^{\star 1}$, Pan-Yang Su$^{1}$, and Shankar Sastry$^{1}$
\thanks{$^\star$Equal contribution.} %
\thanks{Supported by NSF Grant 2031899, Collaborative Research: Transferable, Hierarchical, Expressive, Optimal, Robust, Interpretable Networks.
}
\thanks{$^{1}$Department of Electrical Engineering and Computer Sciences, University of California, Berkeley, CA 94720 (emails: \texttt{\{chihyuan\_chiu, chinmay\_maheshwari, pan\_yang\_su, sastry\} at berkeley dot edu}).}%
}

\begin{document}

\maketitle

\thispagestyle{empty}
\pagestyle{empty}




\begin{abstract}
Arc-based traffic assignment models (TAMs) are a popular framework for modeling traffic network congestion generated by self-interested travelers who sequentially select arcs based on their perceived latency on the network. However, existing arc-based TAMs either assign travelers to cyclic paths, or do not extend to networks with bidirectional arcs (edges)  between nodes. To overcome these difficulties, we propose a new modeling framework for stochastic arc-based TAMs. Given a traffic network with bidirectional arcs, we replicate its arcs and nodes to construct a directed acyclic graph (DAG), which we call the \emph{Condensed DAG} (CoDAG) representation. Self-interested travelers sequentially select arcs on the CoDAG representation to reach their destination. We show that the associated equilibrium flow, which we call the \emph{Condensed DAG equilibrium}, exists, is unique, and can be characterized as a strictly convex optimization problem. 
Moreover, we propose a discrete-time dynamical system that captures a natural adaptation rule employed by self-interested travelers to learn about the emergent congestion on the network. We show that the arc flows generated by this adaptation rule converges to a neighborhood of Condensed DAG equilibrium.  
To our knowledge, our work is the first to study learning and adaptation in an arc-based TAM. Finally, we present numerical results that corroborate our theoretical results. 
\end{abstract}

\section{INTRODUCTION}
\label{sec: Introduction and Related Works}

Traffic assignment models (TAMs) \cite{Cominetti2012Modern,Wardrop1952SomeTheoreticalAspectsOfRoadTrafficResearch,BaillonCominetti2008MarkovianTrafficEquilibrium,fosgerau2013link,daganzo1977stochastic,Akamatsu1997DecompositionOfPathChoiceEntropy,Dial1971AProbabilisticMultipathTrafficAssignmentModel} play a central role in congestion modeling for transportation networks, by informing crucial decisions about infrastructure investment, capacity management, and tolling for congestion regulation. The central dogma behind this modeling approach is that self-interested travelers select routes with minimal \emph{perceived} latency (i.e., the Wardrop or user equilibrium), which can be modeled as deterministic \cite{Cominetti2012Modern,Wardrop1952SomeTheoreticalAspectsOfRoadTrafficResearch} or stochastic \cite{daganzo1977stochastic,Dial1971AProbabilisticMultipathTrafficAssignmentModel,Akamatsu1997DecompositionOfPathChoiceEntropy,BaillonCominetti2008MarkovianTrafficEquilibrium,fosgerau2013link}. Empirical studies confirm that stochastic TAMs achieve greater success at interpreting congestion levels, compared to their deterministic counterparts \cite{sheffi1981comparison}. 

There exist two dominant modeling paradigms in TAM: the route-based model \cite{Cominetti2012Modern, Dial1971AProbabilisticMultipathTrafficAssignmentModel,daganzo1977stochastic,yai1997multinomial}---where each traveler makes a single choice between set of available routes from origin to destination---and the arc (or edge) based model \cite{BaillonCominetti2008MarkovianTrafficEquilibrium, zimmermann2020tutorial, oyama2017discounted, mai2015nested, mai2016method}---where the traveler sequentially makes routing decision at each node on the network, based on their perception of arc latencies. There are two major drawbacks of route-based models on real-world networks: route correlation and route enumeration. Specifically, the utility generated from different routes is correlated due to overlapping arcs on different routes. Moreover, exhaustive route enumeration is prohibitive in terms of computational cost, memory storage, and information acquisition, since the number of routes in a traffic network can be exponential in the number of arcs. 

To avoid explicit route enumeration, Akamatsu \cite{Akamatsu1997DecompositionOfPathChoiceEntropy} proposed the first arc-based stochastic TAM, which was further generalized by Baillon and Cominetti \cite{BaillonCominetti2008MarkovianTrafficEquilibrium}. More recently, Fosgerau et al. and Mai et al. \cite{fosgerau2013link,mai2015nested} presented similar arc-based models based on dynamic discrete choice analysis, which are mathematically similar to the models proposed by Akamatsu \cite{Akamatsu1997DecompositionOfPathChoiceEntropy} and Baillon and Cominetti \cite{BaillonCominetti2008MarkovianTrafficEquilibrium}. However, these models suggest that travelers take cyclic routes with positive probability. To overcome this fundamental modeling challenge, Oyama et al. \cite{oyama2019prism,oyama2022markovian} recently proposed various methods to explicitly avoid routing on cyclic routes.
Unfortunately, these methods either do not apply beyond acyclic graphs \cite{oyama2022markovian} or restrict the set of feasible routes, at the expense of modeling accuracy \cite{oyama2019prism}, or restrictive assumptions on cost structure \cite{BaillonCominetti2008MarkovianTrafficEquilibrium}. Sequential arc selection models in network routing have also been studied by Calderone et al. \cite{Calderone2017MarkovDecisionProcessRoutingGames, calderone2017infinite} 
where each arc selection is accompanied by stochastic transitions to the next arc, and a deterministic transition cost. This stands in contrast to the stochastic TAM literature, where transitions from arc to arc are assumed deterministic and the travel cost (latency) is assumed stochastic. 

In this work, we propose an arc-based stochastic TAM that explicitly avoids cycles by considering routing on a directed acyclic graph derived from the original network, henceforth referred to as the \emph{Condensed Directed Acyclic Graph} (CoDAG). The CoDAG representation duplicates an appropriate subset of nodes and arcs in the original network, to explicitly avoids cycles while preserving all feasible routes. 
Travelers sequentially select arcs on the CoDAG network at every intermediate node, based on perceived arc latencies. 
This route choice behavior is akin to the models prescribed by Akamatsu \cite{Akamatsu1997DecompositionOfPathChoiceEntropy} and Baillon and Cominetti \cite{BaillonCominetti2008MarkovianTrafficEquilibrium}, but with routing occurring over the CoDAG associated with original network. We show that the corresponding equilibrium congestion pattern---which we term the \emph{Condensed DAG equilibrium} (CoDAG equilibrium)---can be characterized as the unique minimizer of a strictly convex optimization problem. 

Moreover, we propose a discrete-time dynamical system that captures a natural adaptation rule used by self-interested travelers who progressively learn towards equilibrium arc selections. In the game theory literature, an equilibrium notion is only considered useful if there exists an adaptive learning scheme that allows self interested players to converge to it \cite{fudenberg1998theory}. 
Despite research progress on both theoretical and algorithmic aspects of stochastic arc-based TAMs, to the best of our knowledge, there has been no research on adaptive learning schemes that ensure convergence to such equilibria. Recently, adaptive learning schemes that converge to equilibria in route-based TAMs have been extensively studied \cite{Krichene2014ConvergenceOfNoRegretLearning,Krichene2015ConvergenceOfHeterogeneousDistributedLearning,Kleinberg2009MultiplicativeUpdates,MaheshwariKulkarni2022DynamicTollingforInducingSociallyOptimalTrafficLoads,Sandholm2010PopulationGamesAndEvolutionaryDynamics,Meigs2017LearningDynamicsinStochasticRoutingGames}, by considering self-interested travelers who repeatedly select routes by observing route latencies in past rounds of interaction. 
In this work, we extend this line of research to arc-based TAMs by proposing a discrete-time dynamics, in which in every round, travelers update arc selections at every node on the CoDAG network based on previous interactions. We prove that the emergent aggregate arc selection probabilities at every node (and the resulting congestion levels on each arc) globally asymptotically converge to a neighborhood of the CoDAG equilibrium. 

To establish convergence, we appeal to the theory of stochastic approximation \cite{Borkar2008StochasticApproximation}, which requires two conditions: (i) The vector field of the discrete-time dynamical system is Lipschitz, and (ii) The trajectories of an associated continuous-time dynamical system asymptotically converge to the CoDAG equilibrium. To prove (i), we establish recursive Lipschitz bounds for vector fields associated with every node. For (ii), we first
construct a Lyapunov function using a strictly convex optimization objective associated with the CoDAG representation. We then show that the value of this Lyapunov function decreases along the trajectories of the continuous-time dynamical system. 
Our contributions are: 
\begin{enumerate}
    \item We introduce a new arc-based traffic equilibrium concept---the \textit{Condensed DAG equilibrium}---which overcomes some limitations of existing 
    traffic equilibrium notions. Furthermore, we show that the Condensed DAG equilibrium is characterized by a solution to a strictly convex optimization problem.
    
    \item We present, to the best of our knowledge, the first adaptive learning scheme in the context of stochastic arc-based TAM. Furthermore, we establish formal convergence guarantees for this learning scheme.
    
    \item We validate our theorems on a simulated traffic network.
\end{enumerate}

The paper proceeds as follows. Section \ref{sec: Preliminaries} introduces the setup considered in this paper, and defines the Condensed DAG representation. Section \ref{sec: New Equilibrium Characterization} defines the Condensed DAG equilibrium, and characterize it as a solution to a strictly convex optimization problem. Section \ref{sec: Methods} presents discrete-time dynamics that converges to the Condensed DAG equilibrium and also provides a proof sketch. In Section \ref{sec: Results}, we numerically study the convergence of the discrete-time dynamics on a simulated traffic network. Finally, Section \ref{sec: Conclusion and Future Work} presents concluding remarks and future work directions.






\paragraph*{Notation} For each positive integer $n \in \N$, we denote $[n] := \{1, \cdots, n\}$. For each $i \in [n]$ in an Euclidean space $\R^n$, we denote by $e_i$ the $i$-th standard unit vector. 

\section{CONDENSED DAG REPRESENTATION}
\label{sec: Preliminaries}

\subsection{Setup}
\label{subsec: Setup}
Consider a traffic network represented by a directed graph $\graphOrig = (\nodesOrig, \arcsOrig)$, possibly with bidirectional arcs, where $\nodesOrig$ and $\arcsOrig$ denote nodes and arcs, respectively. An example is depicted in Figure \ref{fig:Front_Figure___Equivalent_DAG} (top left). 
Let the \textit{origin nodes} and \textit{destination nodes} be two disjoint subsets of nodes in $\graphOrig$. Each traveler enters the network through an origin node to travel to a destination node, by sequentially selecting arcs at every intermediate node. This gives rise to congestion on each arc, which in turn decides the travel times.
Specifically, each arc \(\tilde{a} \in A_O\) is associated with a strictly increasing \emph{latency function} $\latency_{\tilde{a}}: [0, \infty) \ra [0, \infty)$, which gives for each arc the travel time as a function of traffic flow. To simplify our exposition, we assume that there is only one origin-destination tuple \((o,d)\), although the results presented in this paper naturally extend to settings where the traffic network has multiple origin-destination pairs. We denote by \(g_o\) the demand of (infinitesimal) travelers who travel from the origin \(o\) to the destination \(d\).

\begin{remark}
Arc selections made by travelers at different nodes are independent of one another. Therefore, if the underlying network has bidirectional edges, then sequential arc selection by a traveler can result in a cyclic route. 
For example, sequential arc selection in the original network shown on the top left in Figure \ref{fig:Front_Figure___Equivalent_DAG} may lead a traveler to loop between \(i_2^O\) and \(i_3^O\) before reaching destination. 
To overcome this, we introduce a directed acyclic graph (DAG) representation of the original graph \(\graphOrig\) in the following subsections, called the \textit{condensed DAG}. Sequential arc selections made on this network encodes the travel history by design and therefore avoids cyclic routes.
\end{remark}

\subsection{Preliminaries on DAG: Depth and Height}
\label{subsec: Depth and Height}

Before introducing condensed DAG representation, we first present the notions of \textit{height} and \textit{depth} of a DAG. These concepts are crucial for the construction and analysis of condensed DAGs in the following sections.
For the exposition in this subsection, let \(G\) be a DAG with a single origin-destination pair \((o,d)\). Furthermore, let \textbf{R} be the set of all acyclic routes in $\graphMod$ which start at the origin node $o$ and end at the destination node $d$. 


\begin{definition}[\textbf{Depth}] \label{Def: Depth}
For each $r \in \routes$ and $a \in r$, let $\depth_{a,r}$ denote the location of arc $a$ in route $r$, i.e., $a$ is the $\depth_{a,r}$-th arc in the route $r$, and with a slight abuse of notation, define:
\( 
    \depth_a := \max_{r \in \routes: a \in r} \depth_{a,r}, 
\)
We say that $a$ is an \emph{$\depth_a$-th depth arc} in the Condensed DAG $G$.
Moreover, we define the \emph{depth of a node} $i \in \nodesMod\backslash\{o\}$ by:
\begin{align*}
    \depthNode_i := \max_{a \in \arcsMod_i^-} \depth_a
\end{align*}
with \(\depthNode_o = 0\).
\end{definition}

\begin{definition}[\textbf{Height}] \label{Def: Height}
For each $r \in \routes$ and $a \in r$, let $\height_{a,r}$ denote the location of arc $a$ in route $r$ when enumerating arcs in $r$ backwards from the destination node, i.e., $a$ is the $(|r| - \height_{a,r})$-th arc in route $r$, and with a slight abuse of notation, define:
\(
    \height_a := \max_{r \in \routes: a \in r} \height_{a,r}. 
\)
We say that $a$ is an \emph{$\height_a$-th height arc} in the Condensed DAG $\graphMod$
. Moreover, we define the \emph{height of a node} $i \in \nodesMod\backslash\{d\}$ by:
\begin{align*}
    \heightNode_i := \max_{a \in \arcsMod_i^+} \height_a
\end{align*}
with \(\heightNode_d = 0\).
\end{definition}

\subsection{Construction of Condensed DAG}
\label{subsec: Construction of Condensed DAG}

\begin{figure}
\centering
\subfloat{%
  \includegraphics[clip,width=\columnwidth]{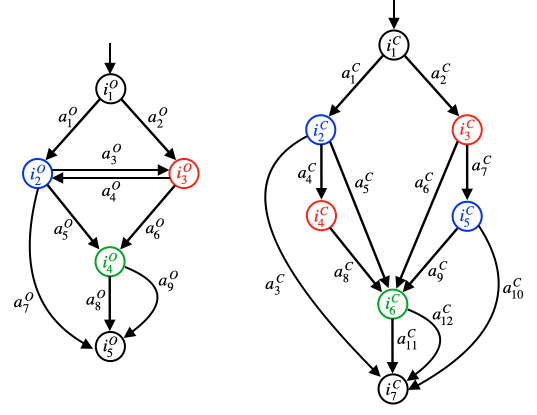}%
}

\subfloat{%
  \includegraphics[clip,width=\columnwidth]{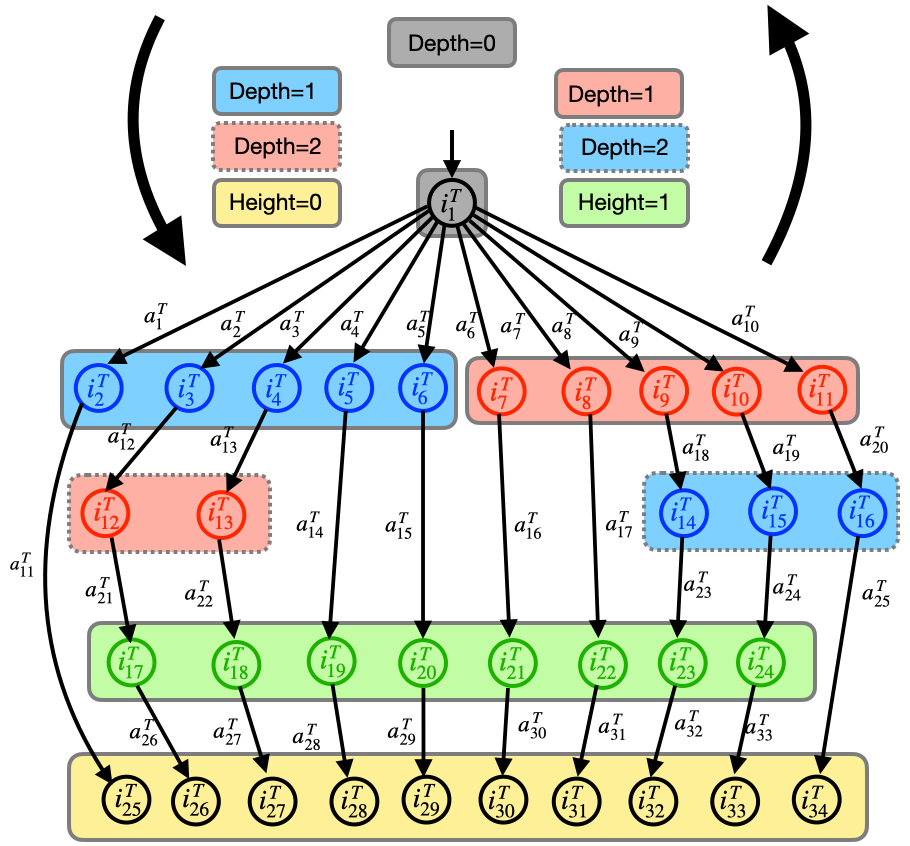}%
}
    \caption{Example of a single-origin single-destination original network $\graphOrig$ (top left, with superscript $O$), and its corresponding tree network (bottom, with superscript $T$) and condensed DAG $\graphMod$ (top right, with superscript $C$). The blocks in \(G_T\) represent a partition \(P_T\) (see \textsf{(S2)}). The depth and height of nodes in every partition are denoted above \(G_T\). Arc correspondences between the three networks are given by Table \ref{table: Equivalent Arcs}, while node correspondences are indicated by color.}
\label{fig:Front_Figure___Equivalent_DAG}
\end{figure}

For ease of description, we illustrate the construction through an example in Figure \ref{fig:Front_Figure___Equivalent_DAG}. We also present a pseudo-code to generate the condensed DAG representation.

A straightforward way to associate $\graphOrig$ with a DAG would be to brute-force enumerate all acyclic (simple) routes and construct a tree network by replicating arcs and nodes by the number of routes passing through them (see Figure \ref{fig:Front_Figure___Equivalent_DAG}, bottom). However, the resulting tree network may contain a significantly larger number of arcs and nodes compared with the original network. To ameliorate this, we present the \textit{condensed DAG} representation (Figure \ref{fig:Front_Figure___Equivalent_DAG}, top right). The condensed DAG is formed by merging superfluous nodes and arcs in the tree network, while ensuring that the graph remains acyclic, and preserving the set of acyclic routes from the original network.

\vspace{1mm}
\renewcommand{\arraystretch}{1.4}
\begin{table}[h!]
\centering
\caption{Arc correspondences between the graphs in Figure \ref{fig:Front_Figure___Equivalent_DAG}: The original network (top left), fully expanded tree (bottom), and the CoDAG (top right).}
\label{table: Equivalent Arcs}
\begin{tabular}{||c || c | c||} 
 \hline
 Original & Tree DAG & CoDAG \\  
 \hline\hline
 \(a_1^O\) & \(a_1^T, a_2^T,a_3^T, a_4^T,a_5^T\) & \(a_1^C\) \\ 
 \hline 
  \(a_2^O\) & \(a_6^T, a_7^T,a_8^T, a_9^T,a_{10}^T\) & \(a_2^C\) \\ 
 \hline 
  \(a_3^O\) & \(a_{12}^T, a_{13}^T\) &  \(a_4^C\)  \\ 
 \hline
  \(a_4^O\) & \(a_{18}^T, a_{19}^T, a_{20}^T\) &  \(a_7^C\)  \\ 
 \hline 
  \(a_5^O\) &\(a_{14}^T, a_{15}^T, a_{23}^T, a_{24}^T\) &  \(a_5^C, a_9^C\)  \\ 
 \hline
  \({a_6^O}\) & \(a_{16}^T, a_{17}^T, a_{21}^T, a_{22}^T\) & \(a_6^C, a_8^C\) \\  
 \hline 
 \(a_7^O\) & \(a_{11}^{T}, a_{25}^T\) & \(a_3^C, a_{10}^C\)\\
 \hline 
\(a_8^O\) & \(a_{26}^T, a_{28}^T, a_{30}^T, a_{32}^T\) & \(a_{11}^C\)\\
 \hline 
 \(a_9^O\) & \(a_{27}^T, a_{29}^T, a_{31}^T, a_{33}^T\) & \(a_{12}^C\)\\
 \hline 
\end{tabular}
\end{table}

One can design a condensed DAG representation as follows: 
\begin{enumerate}   \item[\textsf{(S1)}] Convert the original network \(\graphOrig\) to a tree structure \(\graphMod_T = (I_T, A_T)\), in which every branch emanating from the origin represents a route. Each node and arc is replicated by the number of acyclic routes that contains it. For every node $i$ in \(\graphMod_T\), compute the depth \(\depthNode_i\) and height $\heightNode_i$ (see Definition \ref{Def: Depth}-\ref{Def: Height}).

\item[\textsf{(S2)}] Generate a partition \(P_T\) of \(I_T\) such that:
\begin{itemize}
    \item[(i)] For each \(X\in P_T\), all nodes in \(X\) replicate the node in \(I_O\) that shares the same height or depth in \(G_T\). 
    \item[(ii)] For any \(X,Y\in P_T\), there exists no \(i,i'\in X, j,j'\in Y, \) such that \(\heightNode_j > \heightNode_i\) and \(\heightNode_{j'} < \heightNode_{i'}\) 
\end{itemize}  
\item[\textsf{(S3)}] For each set element $X$ of $P_T$, merge all nodes in $X$ into a single node. Then, merge arcs which have the same start and end nodes, and are replicas of the same edge in the original network \(G_O\).
\end{enumerate}
We refer to any graph generated via \textsf{(S1)}-\textsf{(S3)} as a \emph{condensed DAG} (CoDAG) representation $\graphMod = (\nodesMod, \arcsMod)$ of the original network, where $\nodesMod$ and $\arcsMod$ are the set of nodes and arcs, respectively. By construction, the CoDAG representation explicitly avoids cyclic routes, and preserves all the acyclic routes from the original network. This is because the tree network constructed in \textsf{(S1)} preserves all acyclic routes from original network. Furthermore, the merging conditions stated in \textsf{(S3)} prohibit both the removal and the addition of routes.

\begin{remark}
A given traffic network with bidirectional arcs may yield several distinct CoDAG representations, any of which would be amenable to our analysis in subsequent sections. The development of an algorithmic procedure to compute a CoDAG with the least number of arcs or nodes is beyond the scope of this work.
\end{remark}

\begin{remark}
    The Condensed DAG representation $\graphMod$ can be significantly smaller in size, compared to the tree network. There exist original networks whose corresponding tree representation \(\graphMod_T\) is exponentially larger than its corresponding CoDAG $\graphMod$. For example, consider a network with nodes $i_1, \cdots, i_n$, with two directed arcs connecting $i_k$ to $i_{k+1}$, for each $k \in [n-1]$. Here, the corresponding tree network would have $2^n - 2$ arcs, while the CoDAG representation only has $2(n-1)$ arcs.
\end{remark}

\begin{remark}
The arc-based TAM literature also considers modified representations of traffic networks with bidirectional arcs. For example, Oyama, Hara et al. \cite{oyama2022markovian,papola2013network} consider the Network Generalized Extreme Value (NGEV) representations, which are similar to our CoDAG representation, but applies only to acyclic networks \cite{oyama2022markovian}. Thus, NGEV models cannot capture realistic traffic networks where almost all arcs are bidirectional. Meanwhile, Oyama, Hato et al. \cite{oyama2019prism} consider the Choice Based Prism (CBP) representation, which prunes the available set of feasible routes to ameliorate computational inefficiency. While CBP explicitly avoids cyclic routes, it also removes some acyclic routes during the pruning process. In contrast, the CoDAG representation avoids this issue.
\end{remark}

To conclude this section, we introduce some notation used throughout the rest of the paper. Recall that CoDAGs are formed by replicating the arcs in \(\graphOrig\). To describe this correspondence between arcs, we define $[\cdot]: \arcsMod \ra \arcsOrig$ to be a map from each CoDAG arc $a \in \arcsMod$ to the corresponding arc $[a] \in \arcsOrig$. For each arc $a \in \arcsMod$, let $i_a$ and $j_a$ denote the start and terminal nodes, and for each node $i \in I$, let $\arcsMod_i^-, \arcsMod_i^+ \subset \arcsMod$ denote the set of incoming and outgoing arcs.

\section{EQUILIBRIUM CHARACTERIZATION}
\label{sec: New Equilibrium Characterization}

In this section, we introduce the \emph{condensed DAG (CoDAG) equilibrium} (Definition \ref{Def: CoDAG Equilibrium}),
which is based on the 
CoDAG representation of the original traffic network.  Specifically, we show that the CoDAG equilibrium exists, is unique, and solves a strictly convex optimization problem (Theorem \ref{Prop: Strict Convexity of F}). 

\subsection{Condensed DAG Equilibrium}
\label{subsec: Condensed DAG Equilibrium}

Below, we assume that every traveler knows $G_O$ and has access to the same CoDAG representation of $G_O$. To avoid cyclic routes, we model travelers as performing sequential arc selection over the CoDAG representation $\graphMod = (\nodesMod, \arcsMod)$.
The aggregate effect of the travelers' arc selections gives rise to the congestion on the network. 
Concretely, for each \(a\in A\), let the \textit{flow} or \textit{congestion level} on arc $a$ be denoted by \(w_a\), and let the total flow on the corresponding arc in the original network be denoted, with a slight abuse of notation, by $w_{[a]} := \sum_{a' \in [a]} w_{a'}$. (Note that unlike existing TAMs, the latency of arcs in \(G\) can be coupled through the map \(w_{[\cdot]}\), since multiple copies of the same arc in \(G_O\) may exist in $G$.) Then, the perceived latency of travelers on each arc $a \in A$ is described by:
\begin{align*}
    \tilde s_{[a]}(w_{[a]}) := s_{[a]}(w_{[a]}) + \nu_{a},  
\end{align*}
where $\nu_{a}$ is a zero-mean random variable. At each non-destination node $i \in I \backslash \{d\}$, travelers select among outgoing nodes $a \in A_i^+$ by comparing their perceived latencies-to-go $\tilde z_a: \R^{|A|} \ra \R$, given recursively by:     
\begin{alignat}{2} \label{eq: LTGPerturbed}
    \tilde{z}_a(w) &:= \tilde{s}_{[a]}(w_{[a]}) + \min_{a' \in A_{j_a}^+} \tilde{z}_{a'}(w),  \hspace{5mm} &&j_a \ne d, \\ \nonumber
    \tilde{z}_a(w) &:= \tilde{s}_{[a]}(w_{[a]}), &&j_a = d.
\end{alignat}

Consequently, the fraction of travelers who arrive at \(i\in I\backslash\{d\}\) and choose arc \(a\in A_i^+\) is given by:
\begin{align}\label{eq: Pij}
    P_{ij_a} := \Prob(\tilde z_a \leq \tilde z_{a'}, \hspace{0.5mm} \forall \hspace{0.5mm} a' \in A_i^+).
\end{align}
An explicit formula for the probabilities $\{P_{ij_a}: a \in A_i^+\}$ in terms of the statistics of $\tilde z_a$,
is provided by the discrete-choice theory \cite{BenAkiva1985DiscreteChoiceAnalysis}. In particular, define $z_a(w) := \E[\tilde z_a(w)]$ and $\epsilon_a := \tilde z_a(w) - z_a(w)$, and define the latency-to-go at each node by:
\begin{align} \label{eq: PhiFunc}
    \nodeCostToGo_{i}(\{\costToGo_{a'}(\arcLoadMod): a' \in \arcsMod_i^+ \}) = \E\Bigg[ \min_{a' \in \arcsMod_i^+} 
    \tilde \costToGo_{a'}(\arcLoadMod) 
    \Bigg].
\end{align}
Then, from discrete-choice theory \cite{BenAkiva1985DiscreteChoiceAnalysis}:
\begin{align}\label{eq: ProbTransition}
    P_{ij_a}  = \frac{\partial \varphi_i(z)}{\partial z_a}, \quad i\in I\backslash\{d\}, a\in A_i^{+},
\end{align}
where, with a slight abuse of notation, we write $\nodeCostToGo_i(\costToGo)$ for $\nodeCostToGo_{i}(\{\costToGo_{a'}: a' \in \arcsMod_i^+ \})$. 
To obtain a closed form expression of \(\varphi\), this work considers the \emph{logit Markovian model} \cite{Akamatsu1997DecompositionOfPathChoiceEntropy,BaillonCominetti2008MarkovianTrafficEquilibrium}, which assumes that the zero-mean noise $\epsilon$ is Gumbel-distributed with scale $\beta > 0$. Intuitively, $\beta > 0$ is an entropy parameter that models the degree to which the average traveler's perception of network latency is suboptimal. In this case, the corresponding latency-to-go at each node $i$ in $\graphMod$ is:
\begin{align}\label{eq: PhiFuncLogit}
\nodeCostToGo_i(\costToGo) = - \frac{1}{\beta} \ln\Bigg( \sum_{a' \in \arcsMod_{i}^+} e^{-\beta \costToGo_{a'}} \Bigg).
\end{align}

Using \eqref{eq: LTGPerturbed} and \eqref{eq: PhiFuncLogit}, the expected minimum latency-to-go $\costToGo_a: \R^{|\arcsMod|} \ra \R$,  associated with traveling on each arc $a \in \arcsMod$, is given by:
\begin{align} \label{Eqn: CostToGo}
    \costToGo_a(\arcLoadMod) = \latency_{[a]} \Bigg( \sum_{\bar a \in [a]} \arcLoadMod_{\bar a} \Bigg) - \frac{1}{\beta} \ln\Bigg( \sum_{a' \in \arcsMod_{j_a}^+} e^{-\beta \costToGo_{a'}(\arcLoadMod)} \Bigg).
\end{align}
Note that \eqref{Eqn: CostToGo} is well-posed, as \(z_a\) can be recursively computed along arcs of increasing height (Definition \ref{Def: Height}) from the destination back to the origin. For more details, please see Appendix \ref{subsec: A2, App, Preliminaries} \cite{Chiu2023ArcbasedTrafficAssignment}.

Against the preceding backdrop, we formally define the central equilibrium solution concept studied in this paper: the Condensed DAG Equilibrium (CoDAG Equilibrium).

\begin{definition}[\textbf{Condensed DAG Equilibrium}] \label{Def: CoDAG Equilibrium}
Given $\beta > 0$, a vector of arc-flow $\bar \arcLoadMod^\beta \in \R^{|\arcsMod|}$ is called a \emph{Condensed DAG equilibrium} if, for each $i \in \nodesMod \backslash \{d\}$, $a \in \arcsMod_i^+$:
\begin{align} \label{Eqn: MTE}
   \bar \arcLoadMod_a^\beta &= \left(\nodeLoadIn_{i} + \sum_{a' \in \arcsMod_{i}^+} \bar \arcLoadMod_{a'}^\beta \right) \cdot  \frac{\exp(-\beta \costToGo_a(\bar \arcLoadMod^\beta))}{\sum_{a' \in \arcsMod_{i_a}^+} \exp(-\beta \costToGo_{a'}(\bar \arcLoadMod^\beta))},
\end{align}
where $g_i = g_o$ if $i = o$, $g_i = 0$ otherwise, and $\arcLoadMod \in \arcsLoadConstraintSet$, with:
\begin{align} \label{Eqn: Def, W}
    \arcsLoadConstraintSet := &\Bigg\{ \bar \arcLoadMod^\beta \in \R^{|\arcsMod|}: \sum_{a \in \arcsMod_i^+} \bar \arcLoadMod_a^\beta = \sum_{a \in \arcsMod_i^-} \bar \arcLoadMod_a^\beta, \hspace{0.5mm} \forall \hspace{0.5mm} i \ne o, d, \\ \nonumber
    &\hspace{5mm} \sum_{a \in \arcsMod_o^+} \bar \arcLoadMod_a^\beta = \nodeLoadIn_o, \hspace{1mm} \bar \arcLoadMod_a^\beta \geq 0, \hspace{0.5mm} \forall \hspace{0.5mm} a \in \arcsMod \Bigg\}.
\end{align}
\end{definition}

\vspace{1mm}
For any CoDAG equilibrium \(\bar{w}^{\beta}\), the fraction of travelers at any node \(i\in I\backslash\{d\}\) who selects an arc \(a\in A_i^+\) is:
\begin{align*}
    \bar \probDist_a^\beta := \frac{\bar \arcLoadMod_a^\beta}{\sum_{a' \in A_{i}^+} \bar \arcLoadMod_{a'}^\beta}.
\end{align*}

\begin{remark}
Essentially, at the CoDAG equilibrium, the traveler population at each intermediate node $i \in \nodesMod \backslash \{d\}$ (with total flow $g_i + \sum_{a' \in \arcsMod} \arcLoadMod_{a'}$) select from outgoing arcs by comparing their costs-to-go using the softmax function. While the CoDAG equilibrium and Markovian Traffic Equilibrium (MTE) share some similarities (see \cite{BaillonCominetti2008MarkovianTrafficEquilibrium}), there also exist two main fundamental differences. First, by design, the CoDAG equilibrium does not yield cyclic routes with strictly positive probability (as is the case with the MTE).
Second, unlike the MTE, congestion levels on arcs (which may be replicas of the same arc in \(G_O\)) in the CoDAG representation are coupled to each other. Therefore, MTE analysis does not extend straightforwardly to the CoDAG equilibrium.
\end{remark}

\subsection{Existence and Uniqueness of the CoDAG equilibrium}

In this subsection, we show the existence and uniqueness of the CoDAG equilibrium, by characterizing it as the unique minimizer of a strictly convex optimization problem over a compact set. First, for each $[a]\in A_O$, define:
\begin{align}
    f_{[a]}(\arcLoadMod) := \int_0^{\arcLoadMod_{[a]}}  \latency_{[a]}(u) \hspace{0.5mm} du,
\end{align}
and for each \(i\in I\backslash\{d\}\), set:
\begin{align}
    \chi_i(\arcLoadMod_{\arcsMod_i^+}) := \sum_{a \in \arcsMod_i^+} \arcLoadMod_a \ln \arcLoadMod_a - \Bigg(\sum_{a \in \arcsMod_i^+} \arcLoadMod_a \Bigg) \ln \Bigg(\sum_{a \in \arcsMod_i^+} \arcLoadMod_a \Bigg).
\end{align}
Finally, define $F: \arcsLoadConstraintSet \ra \R$ by:
\begin{align} \label{Eqn: Def, F}
    F(w) &= \sum_{[a]\in A_O}f_{[a]}(w) + \frac{1}{\beta}\sum_{i\neq d}\chi_i(w_{A_i^+}),
\end{align}
where $\arcLoadMod_{\arcsMod_i^+} \in \R^{|\arcsMod_i^+|}$ denotes the components of $\arcLoadMod$ corresponding to arcs in $\arcsMod_i^+$.

\begin{theorem} \label{Thm: CoDAG is unique minimizer of F}
The CoDAG equilibrium $\bar \arcLoadMod^\beta \in \arcsLoadConstraintSet$ exists, is unique, and is the unique minimizer of $F$ over $\arcsLoadConstraintSet$.
\end{theorem}



To prove Theorem \ref{Thm: CoDAG is unique minimizer of F}, we first show that \(F(\cdot)\) is strictly convex over $\arcsLoadConstraintSet$ (Lemma \ref{Prop: Strict Convexity of F}), so $F$ has a unique minimizer in $\arcsLoadConstraintSet$. It then suffices to show that the CoDAG equilibrium definition (Definition \ref{Def: CoDAG Equilibrium}) matches the Karush-Kuhn-Tucker (KKT) conditions for the optimization problem \eqref{Eqn: Def, F}.  


\begin{lemma} \label{Prop: Strict Convexity of F}
The map $F: \arcsLoadConstraintSet \ra \R$ is strictly convex.
\end{lemma}

\begin{proof}(\textbf{Proof Sketch})
It suffices to show that $f_{[a]}$ and $\chi_i$ are convex for each $[a] \in \arcsOrig$, $i \in \nodesMod \backslash \{d\}$. Each $f_{[a]}$ is convex, since it is the composition of a convex function (\(w\mapsto \sum_{a\in A_O}\int_{0}^{w_a}s_a(u) d u\)) with a linear function ($\arcLoadMod_{[a]} := \sum_{a' \in [a]} \arcLoadMod_{a'}$). Furthermore, we establish that for any \(i\in I \backslash\{d\}\), $y_i \in \R^{|\arcsMod_i^+|}$:
\begin{align*} \nonumber
    y_i^\top \nabla_\arcLoadMod^2 \chi_i(\arcLoadMod) y_i \geq 0,
\end{align*}
where the equality holds if and only if \(y_i\) and \(w_{A_i^+}\) are scalar multiples of one another. Strict convexity then follows by a contradiction argument showing that there exists at least one node \(i\in I\backslash\{d\}\) such that \(y_i^\top \nabla_\arcLoadMod^2 \chi_i(\arcLoadMod) y_i>0\).
\end{proof}



\section{LEARNING DYNAMICS}
\label{sec: Methods}
In this section, we propose a discrete-time dynamical system \eqref{Eqn: General Network, xi flow, discrete}
which captures travelers' preferences for minimizing total travel time, as well as their perception uncertainties, while simultaneously learning about the emergent congestion on the network. 

We leverage the constant step-size stochastic approximation theory to prove that these discrete-time dynamics converge to a neighborhood of the CoDAG equilibrium (Theorem \ref{Thm: Convergence, w, discrete}). To this end, we first prove that the continuous-time counterpart to \eqref{Eqn: General Network, xi flow, discrete} globally asymptotically converges to the CoDAG equilibrium (Lemma \ref{Lemma: Convergence, w, continuous}).
We then conclude the proof by verifying technical assumptions required to invoke results in stochastic approximation theory \cite{Borkar2008StochasticApproximation} (Lemma \ref{Lemma: Technical Conditions for Stochastic Approximation}).


\subsection{Discrete-time Dynamics}
In this subsection, we present a discrete-time dynamical equation that captures the evolution of flows on the network as a result of learning and adaptation by self-interested travelers. More formally,
at each discrete time step $n \geq 0$, \(g_o\) units of travelers arrive at the origin node \(o\). At time step \(n\), every traveler who reaches node \(i\in I \backslash\{d\}\) selects some arc \(a\in A_i^+\). For any \(i\in I \backslash\{d\}, a\in A_i^+\), let \(\probDist_a[n]\) be the \emph{aggregate arc selection probability}: the fraction of travelers at node \(i\) choosing arc \(a\) at time \(n\).
As a result of the arc selections made by every traveler, a flow of \(W[n]\) is induced on the arcs as given below. For every \(a\in A\):
\begin{align} \label{Eqn: General Network, w flow, discrete}
\arcLoadModDiscrete_a[n] &= \Bigg( g_{i_a} + \sum_{a' \in \arcsMod_{i_a}^-} \arcLoadModDiscrete_{a'}[n] \Bigg) \cdot \probDist_a[n],
\end{align}
where, as given in Definition \ref{Def: CoDAG Equilibrium}, $g_{i_a} = g_o$ if $i_a = o$, and $g_{i_a} = 0$ otherwise.


At the end of each time step, every traveler reaches the destination and observes a noisy estimate of the latency-to-go, independent across travelers, on every arc in the network (including ones they did not visit in that time step). Note that the latency-to-go for any arc is dependent on the congestion \(W[n]\), which in turn depends on aggregate decisions taken by travelers (please refer to \eqref{Eqn: General Network, w flow, discrete}). 
Based on the observed latencies, at time \(n+1\), at every non-destination node \(i\in I\backslash\{d\}\), a \(\eta_{i}[n+1] \cdot K_i\) fraction of travelers at node \(i\) switches to an arc with the minimum observed latency-to-go. Meanwhile, a \(1-\eta_{i}[n+1] \cdot K_i\) fraction of travelers selects the same arc they selected at time step \(n\). We assume that $\{\eta_i[n+1] \in \R: i \in \nodesMod, n \geq 0 \}$ are independent bounded random variables
in $[\underline \mu,  \overline \mu]$, independent of travelers' perception stochasticities, with 
$0 < \underline \mu < \mu < \overline \mu < 1/\max\{K_i: i \in \nodesMod \backslash \{d\}\}$ and 
$\E[\eta_{i_a}[n+1]] = \mu$ for each node index $i \in I$ and discrete time index $n \geq 0$. Meanwhile, the constants $K_i$ represent node-dependent update rates. To summarize, the dynamic evolution of arc selections by infinitesimal travelers is captured by the following evolution of \(\probDist[n]\). For every \(i\in I\backslash\{d\}, a\in A_i^+\):
\begin{align*}
    &\probDist_a[n+1] 
    = \probDist_a[n] + \eta_{i_a}[n+1] \cdot K_{i_a} \left(-\probDist_a[n] + P_{ij_a} \right),
\end{align*}
where \(P_{ij_a}\) is defined in \eqref{eq: Pij}. Using \eqref{eq: ProbTransition} and \eqref{eq: PhiFuncLogit}, the previous equation can be rewritten as: 
\begin{align} \label{Eqn: General Network, xi flow, discrete} \tag{\textsf{PBR}}
    &\probDist_a[n+1] \\ \nonumber
    = \hspace{0.5mm} &\probDist_a[n] + \eta_{i_a}[n+1] \cdot K_{i_a} \\ \nonumber
    &\hspace{5mm} \cdot \Bigg(-\probDist_a[n] + \frac{\exp(-\beta \big[ \costToGo_a(\arcLoadModDiscrete[n]) \big])}{\sum_{ a' \in \arcsMod_{i_a}^+} \exp(-\beta \big[ \costToGo_{a'}(\arcLoadModDiscrete[n]) \big])} \Bigg),
\end{align}
The dynamics \eqref{Eqn: General Network, xi flow, discrete} bears close resemblance to perturbed best response dynamics in routing games \cite{Sandholm2010PopulationGamesAndEvolutionaryDynamics}, so we shall refer to \eqref{Eqn: General Network, xi flow, discrete} as \emph{perturbed best response} dynamics.

We assume $\probDist_a[0] > 0$ for each $a \in \arcsMod$, i.e., each arc has some strictly positive initial traffic flow. This is reasonable, since the stochasticity in travelers' perception of network congestion ensures that each arc has a nonzero probability of being selected.

\subsection{Convergence Results}
Our main theorem establishes that the discrete-time dynamics 
\eqref{Eqn: General Network, xi flow, discrete}  asymptotically converges to a neighborhood of the CoDAG equilibrium $\bar \arcLoadMod^\beta$.

\begin{theorem} \label{Thm: Convergence, w, discrete}
Under the discrete-time flow dynamics \eqref{Eqn: General Network, xi flow, discrete}, for each $\delta > 0$:
\begin{align*}
    \limsup_{n \ra \infty} \E\big[ \Vert \probDist[n] - \bar \probDist^\beta \Vert_2^2 \big] &\leq O(\mu), \\
    \limsup_{n \ra \infty} \Prob\big( \Vert \probDist[n] - \bar \probDist^\beta \Vert_2^2 \geq \delta \big) &\leq O\left( \frac{\mu}{\delta} \right).
\end{align*}
\end{theorem}

\vspace{2mm}
To prove Theorem \ref{Thm: Convergence, w, discrete}, we leverage the theory of constant step-size stochastic approximation \cite{Borkar2008StochasticApproximation}. This requires proving that the continuous-time dynamics corresponding to the discrete-time update \eqref{Eqn: General Network, xi flow, discrete}, presented below, converges to the CoDAG equilibrium. For each arc $a \in \arcsMod$:
\begin{align} \label{Eqn: General Network, xi flow, continuous}
    \dot \probDist_a(t) &= - K_i \left( \probDist_a(t) + \frac{\exp(-\beta \cdot \costToGo_a(\arcLoadMod(t)))}{\sum_{ a' \in \arcsMod_{i_a}^+} \exp(-\beta \cdot \costToGo_{a'}(\arcLoadMod(t)))} \right),
\end{align}
where $\arcLoadMod(t)$ is the resulting arc flow associated with the arc selection probability $\probDist(t)$, similar to \eqref{Eqn: General Network, w flow, discrete}:
\begin{align} \label{Eqn: General Network, w flow, continuous}
    \arcLoadMod_a(t) &= \probDist_a(t) \cdot \Bigg(\nodeLoadIn_{i_a} + \sum_{a' \in \arcsMod_{i_a}^-} \arcLoadMod_{a'}(t)\Bigg).   
\end{align}

\begin{lemma}[\textbf{Informal}]
\label{Lemma: Convergence, w, continuous}
Suppose $\arcLoadMod(0) \in \arcsLoadConstraintSet$, i.e., the initial flow satisfies flow continuity. Under the continuous-time flow dynamics \eqref{Eqn: General Network, w flow, continuous} and \eqref{Eqn: General Network, xi flow, continuous}, if $K_i \ll K_{i'}$ whenever $\ell_i < \ell_{i'}$, the traffic flow $\arcLoadMod(t)$ globally asymptotically converges to the CoDAG equilibrium $\bar \arcLoadMod^\beta$.
\end{lemma}

\begin{proof}(\textbf{Proof Sketch})
Recall that $\bar \arcLoadMod^\beta$ is the unique minimizer of the map $F: \arcsLoadConstraintSet \ra \R$, defined by \eqref{Eqn: Def, F}. We show that $F$ is a Lyapunov function for the continuous-time flow dynamics \eqref{Eqn: w flow, recursive, with h} induced by the arc selection dynamics \eqref{Eqn: General Network, xi flow, continuous}. To this end, we first unwind the dynamics \eqref{Eqn: General Network, xi flow, continuous} and \eqref{Eqn: General Network, w flow, continuous} to obtain the recursive relation:
{\small
\begin{align*}
    \dot\arcLoadMod_a(t) &= - K_{i_a} \Bigg( 1 - \frac{1}{K_{i_a}} \cdot \frac{ \sum_{a' \in \arcsMod_{i_a}^-} \dot \arcLoadMod_{a'}(t)}{\sum_{\hat a \in \arcsMod_{i_a}^+} \arcLoadMod_{\hat a}(t)} \Bigg) \arcLoadMod_a(t) \\ \nonumber
    &\hspace{5mm} + K_{i_a} \cdot \sum_{a' \in \arcsMod_{i_a}^-} \arcLoadMod_{a'}(t) \cdot \frac{\exp(-\beta \costToGo_a(\arcLoadMod(t)))}{\sum_{a' \in \arcsMod_{i_a}^+} \exp(-\beta \costToGo_{a'}(\arcLoadMod(t)))}.
\end{align*}
}
Then, we establish that along any trajectory starting on $\arcsLoadConstraintSet$ and following the dynamics given by \eqref{Eqn: General Network, xi flow, continuous}, we have for each $t \geq 0$:
\begin{align*}
    \dot F(t) &= \dot w(t)^\top \nabla_w F(w(t)) \leq 0.
\end{align*}
The proof then follows from LaSalle's Theorem (see \cite[Proposition 5.22]{Sastry1999NonlinearSystems}). For a precise characterization and detailed proof of Lemma \ref{Lemma: Convergence, w, continuous}, please see Appendix \ref{subsec: A3, Methods} \cite{Chiu2023ArcbasedTrafficAssignment}.
\end{proof}

\begin{remark}
On a technical level, the statement and proof technique of Theorem \ref{Thm: Convergence, w, discrete} share similarities with methods used to establish the convergence of best-response dynamics in potential games
\cite{Sandholm2010PopulationGamesAndEvolutionaryDynamics}. However, there exist crucial distinctions between the two approaches which render our problem more difficult. First, since the map $F$ defined by \eqref{Eqn: Def, F} is not a potential function, the mathematical machinery of potential games cannot be directly applied. Moreover, the continuous-time flow dynamics \eqref{Eqn: General Network, xi flow, continuous} and \eqref{Eqn: General Network, w flow, continuous} allow couplings between arbitrary arcs in the CoDAG. For more details, please see Appendix \ref{subsec: A3, Methods} \cite{Chiu2023ArcbasedTrafficAssignment}.
\end{remark}

\begin{remark}
The assumption that $K_i \ll K_{i'}$ whenever the depth of node $i \in \nodesMod \backslash \{d\}$ is less than the depth of node $i' \in \nodesMod \backslash \{d\}$ conforms to the intuition that travelers farther away from the destination node face more complex route selection decisions based on more information regarding traffic flow throughout the rest of the network, and thus perform slower updates.
\end{remark}

Having established the global asymptotic convergence of the continuous-time dynamics \eqref{Eqn: General Network, xi flow, continuous} and \eqref{Eqn: General Network, w flow, continuous} to the CoDAG equilibrium $\bar \arcLoadMod^\beta$, it remains to verify the remaining technical conditions necessary to prove Theorem \ref{Thm: Convergence, w, discrete} via stochastic approximation theory. To this end, we rewrite the discrete $\xi$-dynamics \eqref{Eqn: General Network, xi flow, discrete} as a Markov process with a martingale difference term:
\begin{align*}
    \probDist_a[n+1] &= \probDist_a[n] + \mu \big(\rho_a(\xi[n]) + M_a[n+1] \big),
\end{align*}
where $\rho_a: \R^{|\arcsMod|} \times \R^{|\arcsOrig|} \ra \R^{|\arcsMod|}$ is given by:
\begin{align} \label{Eqn: rho a}
    \rho_a(\probDist) &:= K_{i_a} \left( -\probDist_a + \frac{\exp(-\beta \cdot  \costToGo_a(\arcLoadMod))}{\sum_{ a' \in \arcsMod_{i_a}^+} \exp(-\beta \cdot \costToGo_{a'}(\arcLoadMod))} \right),
\end{align}
with $\arcLoadMod \in \R^{|\arcsMod|}$ defined arc-wise by $\arcLoadMod_a = (g_{i_a} + \sum_{\hat a \in A_{i_a}^-} w_{a'}) \cdot \probDist_a$, and:
\begin{align} 
\label{Eqn: Ma, Discrete-Time Dynamics}
    M_a[n+1] 
    := \hspace{0.5mm} &\left( \frac{1}{\mu} \eta_{i_a}[n+1] - 1 \right) \cdot \rho_a(\xi[n]).
\end{align}
Here, $\arcLoadModDiscrete_a[n] = \big( \nodeLoadIn_{i_a} + \sum_{a' \in \arcsMod_{i_a}^-} W_{a'}[n] \big)$, as given by \eqref{Eqn: General Network, w flow, discrete}.

The following lemma bounds the magnitude of the discrete-time flow $\arcLoadModDiscrete[n] \in \R^{|\arcsMod|}$ and the martingale difference terms $M[n] \in \R^{|\arcsMod|}$.

\begin{lemma} \label{Lemma: Technical Conditions for Stochastic Approximation}
Given initial flows $W[0]$ and arc selection probabilities $\xi[0]$:
\begin{enumerate}
    \item For each $a \in \arcsMod$: $\{M_a[n+1]: n \geq 0\}$ is a martingale difference sequence with respect to the filtration $\mathcal{F}_n := \sigma\big( \cup_{a \in \arcsMod} (\arcLoadModDiscrete_a[1], \probDist[1], \cdots, \arcLoadModDiscrete_a[n], \probDist[n]) \big)$.
    
    \item There exist $C_w, C_m > 0$ such that, for each $a \in \arcsMod$, $n \geq 0$, we have $\arcLoadModDiscrete_a[n] \in [C_w, g_o]$ and $|M_a[n]| \leq C_m$.
    
    \item For each $a \in \arcsMod$, the map $\rho_a$, given by \eqref{Eqn: rho a}, is Lipschitz continuous over the range of realizable flow and arc selection probability trajectories $\{W[n]: n \geq 0\}$ and $\{\xi[n]: n \geq 0\}$.
\end{enumerate}
\end{lemma}

\begin{proof}(\textbf{Proof Sketch})
The first part of Lemma \ref{Lemma: Technical Conditions for Stochastic Approximation} follows because, with respect to $\mathcal{F}_n$, the only stochasticity in $M_a[n+1]$ originates from the i.i.d. input flows $\eta_{i_a}[n+1]$. The second part follows by invoking the flow continuity equations in \eqref{Eqn: General Network, w flow, discrete} to recursively upper bound each $\arcLoadModDiscrete_a[n]$ and $\costToGo_a(\arcLoadModDiscrete[n])$, in increasing order of depth and height, respectively (flows are propagated from origin to destination, and latency-to-go values are computed in the opposite direction). These bounds are then used to recursively establish upper and lower bounds for each $\xi_a[n]$, and consequently each $\arcLoadModDiscrete[n]$, in order of increasing depth. Finally, the Lipschitz continuity of each $\rho_a$ can be proved by establishing that $\rho_a$ is continuously differentiable, with bounded derivatives over the compact domain defined by the bounds on $\arcLoadModDiscrete[n]$ established in the second part of the lemma. For details, please see the proofs of Lemmas \ref{Lemma: Boundedness of Maps} and \ref{Lemma: Lipschitz Continuity of Maps} in Appendix \ref{subsec: A3, Methods} \cite{Chiu2023ArcbasedTrafficAssignment}. 
\end{proof}

\section{EXPERIMENT RESULTS}
\label{sec: Results}

\begin{table}
\centering\footnotesize
\caption{Parameters for simulation.}
\label{table: Parameters for simulation}\def\arraystretch{.9}
\begin{tabular}[t]{p{1cm}p{5.5cm}}
\toprule
{\bf Notation} & {\bf Default value}\\
\midrule
$k_0$ & 0, 1, 0, 1, 1, 0, 1, 1, 1 (ordered by edge index)\\
$k_1$ & 2, 1, 1, 1, 1, 1, 2, 2, 2 (ordered by edge index) \\
$g_1$ & 1\\
$\beta$ & 10\\
$\eta_{i_a}[n]$ & Uniform($0, 0.1$), $\forall a \in \arcsMod, i \in \nodesMod \backslash \{d\}$\\
\bottomrule
\end{tabular}
\end{table}

\begin{figure}
    \centering
    \includegraphics[scale=0.22]{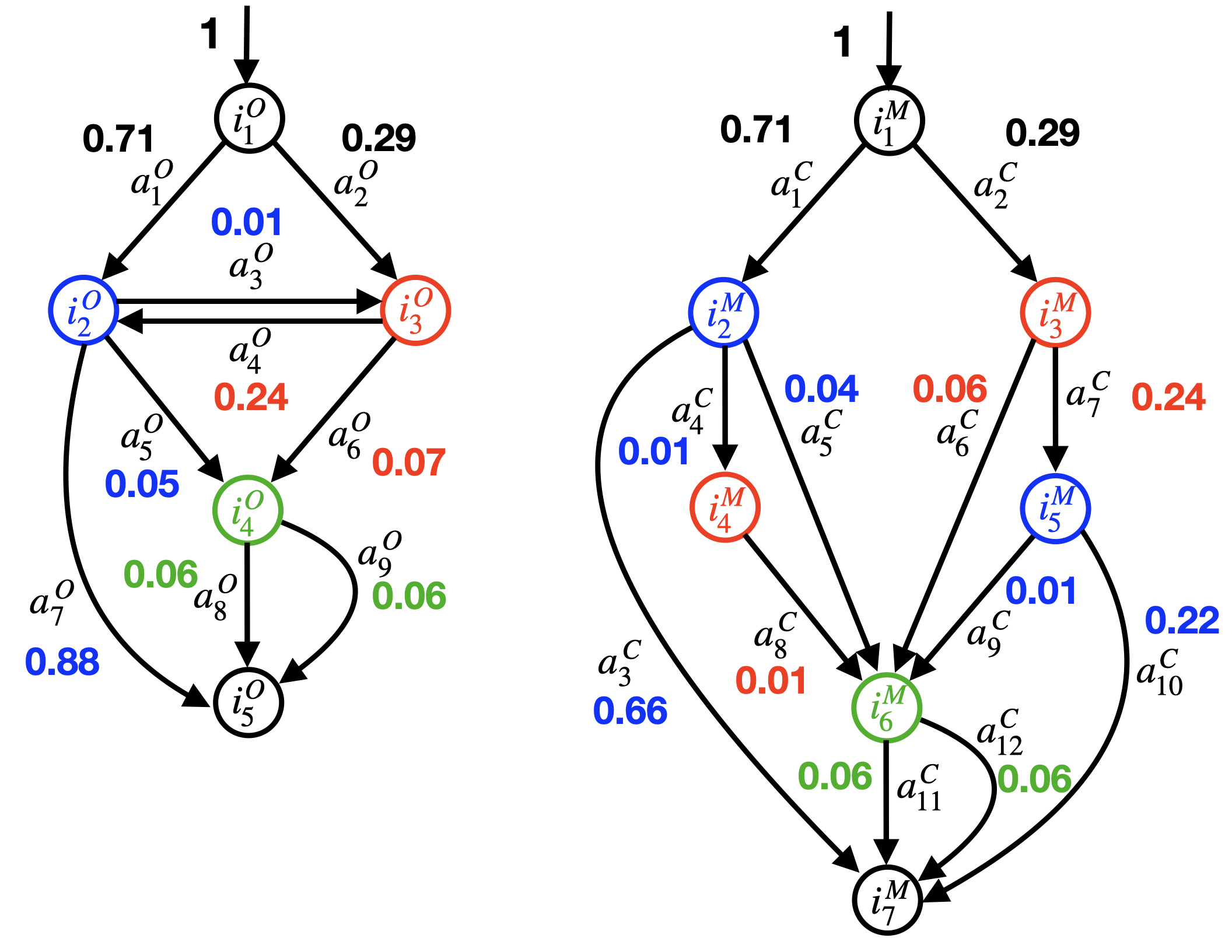}
    \caption{Steady state traffic flow on each arc for an original network and condensed DAG. Flows on arcs emerging from same node are represented in same color.}
\label{fig:finalVer2}
\end{figure}

In this section, we conduct numerical experiments to validate the theoretical analysis presented in Section \ref{sec: Methods}. 
We show in simulation that, under \eqref{Eqn: General Network, xi flow, discrete}, the traffic flows converge to a neighborhood of the condensed DAG equilibrium, as claimed by Theorem \ref{Thm: Convergence, w, discrete}.

\begin{figure}
    \centering
    \includegraphics[scale=0.35]{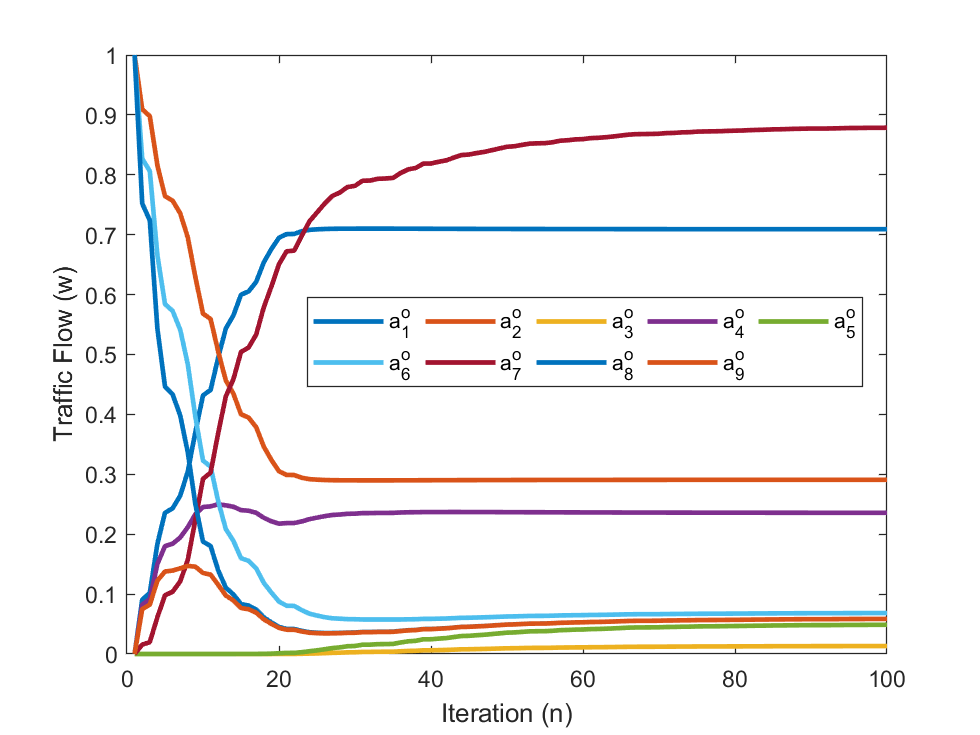}
    \caption{Traffic flow $W[n]$ for the network in Fig. \ref{fig:finalVer2}.}
    \label{fig:w_dynamics}
\end{figure}

Consider the network presented in Figure \ref{fig:Front_Figure___Equivalent_DAG}, with affine edge-latency functions $\latency_{[a]}(w_{[a]}) = k_0 + k_1 \arcLoadMod_{[a]}$ for each arc $a \in A$, where $k_0, k_1 > 0$ are simulation parameters provided in Table \ref{table: Parameters for simulation}. To validate Theorem \ref{Thm: Convergence, w, discrete}, we evaluate and plot the traffic flow values $W_a[n]$ on each arc $a \in \arcsMod$ and discrete time $n \geq 0$. 
Figure \ref{fig:finalVer2} presents traffic flow values at the condensed DAG equilibrium (i.e., $\arcLoadMod^\beta$) for the original network and condensed DAG. While travelers generally prefer routes of lower latency, each route has a nonzero level of traffic flow at equilibrium. The reason is that under the perturbed best response dynamics, users do not allocate all the traffic flow to the minimum-cost route, but instead distribute their traffic allocation more evenly. 
Meanwhile, Figure \ref{fig:w_dynamics} illustrates that $w$ converges to the condensed DAG equilibrium in approximately 100 iterations with some initial fluctuations. The fluctuations are due to the magnitude of the average step-size $\mu$. If $\mu$ is small, the discrete-time update is close to the continuous-time dynamics, and the resulting evolution of the traffic flow follows a smoother trend. Note that in practice, flow convergence to the CoDAG equilibrium occurs even when the effects of the constants $\{K_i: i \in \nodesMod \}$ are ignored, i.e., when each $K_i$ is set to unity.


\section{CONCLUSION AND FUTURE WORK}
\label{sec: Conclusion and Future Work}

We present a new equilibrium concept for stochastic arc-based TAMs in which travelers are guaranteed to be routed on acyclic routes. Specifically, we construct a condensed DAG representation of the original network, by replicating arcs and nodes to avoid cyclic routes, while preserving the set of feasible routes from the original network. We characterize the proposed equilibrium as the optimal solution of a strictly convex optimization problem. Furthermore, we propose adaptive learning dynamics for arc-based TAM that characterizes the evolution of flow generated by the simultaneous learning and adaptation of self-interested travelers. Additionally, we prove that the learning dynamics converges to the corresponding equilibrium flow allocation.

Interesting avenues of future research include: (i) Developing an equilibrium notion and corresponding convergent learning dynamics, for the case in which travelers can only access latency-to-go values on the routes they choose; and (ii) Developing dynamic tolling mechanisms to properly allocate equilibrium flows to induce socially optimal loads.


\printbibliography

\appendix

Please use the following link to access an ArXiv version with the appendix \cite{Chiu2023ArcbasedTrafficAssignment} (\url{https://arxiv.org/pdf/2304.04705.pdf}). 
 

\newpage

Below, we present proofs omitted in the main paper due to space limitations.

\subsection{Properties of Depth and Height}
\label{subsec: A1, Properties of Depth, Height}

In the main text, we recursively defined some dynamical quantities, such as the time evolution of the traffic flows $\arcLoadMod \in \R^{|A|}$ and the latency-to-go $z \in \R^{|A|}$, in a component-wise fashion, either from the origin of the Condensed DAG $\graphMod$ towards the destination, or from the destination to the origin. To facilitate these recursive definitions, we require the following characterizations regarding the depths and heights of arcs in a Condensed DAG $G$.

\subsubsection{Depth}
\label{subsubsec: A1, Depth}

First, we define the concept of depth of a directed acyclic graph (DAG), which will be crucial for the remaining exposition.

\begin{definition}[\textbf{Depth of a DAG}]
Given a DAG $\graphMod = (\nodesMod, \arcsMod)$ describing a single-origin single-destination traffic network, the \textit{depth of $\graphMod$}, denoted $\depth(G)$, is defined by: 
\begin{align*}
    \depth(\graphMod) := \max_{a \in \arcsMod} \depth_a.
\end{align*}
\end{definition}

\vspace{2mm}
In this work, we consider only acyclic routes in traffic networks with finitely many edges, so we have $\depth(\graphMod) < \infty$. Moreover, the case $\depth(\graphMod) = 1$ corresponds to a parallel link network, for which the results of the following proposition have already been analyzed in \cite{MaheshwariKulkarni2022DynamicTollingforInducingSociallyOptimalTrafficLoads}. Therefore, we assume below that $\depth(G) \geq 2$.

\begin{proposition} \label{Prop: Condensed DAG Properties, Depth}
Given a Condensed DAG $\graphMod = (\nodesMod, \arcsMod)$ with the route set $\routes$:
\begin{enumerate}
    \item For any $a \in \arcsMod$, we have $\depth_a = 1$ if and only if $i_a = o$. Similarly, if $\depth_a = \depth(\graphMod)$, then $j_a = d$.
    
    \item For any fixed $r \in \routes$, and any $a, a' \in r$ with $\depth_{a, r} < \depth_{a',r}$, we have $\depth_a < \depth_{a'}$ i.e., arcs along a route have strictly increasing depth from the origin to the destination.
    
    \item Fix any $a \in \arcsMod$, and any $r \in \routes$ containing $a$ such that $\depth_{a,r} = \depth_a$. Then, for any $a' \in \routes$ preceding $a$ in $r$, we have $\depth_{a',r} = \depth_{a'}$.
    
    \item For each depth $k \in [\depth(G)] := \{1, \cdots, \depth(G)\}$, there exists some $a \in A$ such that $\depth_a = k$.
\end{enumerate}
\end{proposition}

\begin{proof} ~\
\begin{enumerate}
    \item If $\depth_a \ne 1$, then $\depth_a \geq 2$, so there exists at least one route $r \in \routes$ containing $a \in A$ such that $\depth_{a,r} \geq 2$. Thus, $i_a \ne o$ (otherwise the first $\depth_{a,r}-1$ arcs of $r$ would form a cycle). Conversely, if $i_a \ne o$, then no route $r \in \routes$ contains $a \in A$ as its first arc, i.e., $\depth_{a,r} \geq 2$ for each $r \in \routes$ containing $a$. Thus, $\depth_a = \max_{r \in \routes: a \in r} \depth_{a,r} \geq 2$; in particular, $\depth_a \ne 1$. This establishes that $\depth_a = 1$ if and only if $i_a = o$.
    
    $\hspace{5mm}$ Now, suppose by contradiction that there exists some $a \in A$ such that $\depth_a = \depth(G)$ but $j_a \ne d$. Fix any $r \in \routes$ such that $a \in r$ and $\depth_{a,r} = \depth_a$. Then $a$ cannot be at the end of $\routes$, since by definition, routes must end at $d$. Let $a' \in r$ be the arc immediately after $a$ in $r$. Then $\depth_{a'} \geq \depth_{a',r} = \depth_{a,r} + 1 = \depth(G) + 1$, a contradiction to the definition of $\depth(G)$.
 
    \item Fix $r \in \routes$, $a,a' \in r$ such that $\depth_{a,r} < \depth_{a',r}$. If $\depth_a = 1$, then $\depth_{a'} \geq \depth_{a',r} > \depth_{a, r} = 1 = \depth_a$, and we are done. Suppose $\depth_a \geq 2$. By definition of $\depth_a$, there exists some route $r_2$ such that $\depth_{a, r_2} = \depth_a$. Construct a new route $r_3 \in \routes$ by replacing the first $\depth_{a,r}$ arcs of $r$ with the first $\depth_{a,r_2}$ arcs of $r_2$. Then $\depth_{a'} \geq \depth_{a', r_3} = \depth_{a', r} - \depth_{a, r} + \depth_{a, r_2} > \depth_{a, r_2} = \depth_a$.
    
    \item Fix any $a \in A$, and any $r \in \routes$ containing $a$ such that $\depth_{a,r} = \depth_a$. Suppose by contradiction that there exists some $a' \in \routes$, preceding $a$ in $r$, for which $\depth_{a'} \geq \depth_{a',r} + 1$. Then, by applying the second part of this lemma along the $(\depth_{a,r} - \depth_{a', r})$ arcs of $\routes$ from $a'$ to $a$, we find that $\depth_a \geq \depth_{a'} + (\depth_{a,r} - \depth_{a', r}) \geq \depth_{a,r} + 1 = \depth_a + 1$, a contradiction.
    
    \item Fix any arc $a \in A$ with $\depth_a = \depth(G)$. Then there exists some $r \in \routes$ containing $a$ such that $\depth_{a,r} = \depth_a = \depth(G)$. It follows from the third part of this proposition that, for each $k \in [\depth(G)]$, the $k$-th arc in $\routes$ is of depth $k$.
    
\end{enumerate}
\end{proof}

\subsubsection{Height}
\label{subsubsec: A1, Height}


Next, we define the concept of height of a directed acyclic graph (DAG), which will be crucial for the remaining exposition.

\begin{definition}[\textbf{Height of a DAG}]
Given a DAG $\graphMod = (\nodesMod, \arcsMod)$ describing a single-origin single-destination traffic network, the \textit{height of $\graphMod$}, denoted $\height(G)$, is defined by: 
\begin{align*}
    \height(\graphMod) := \max_{a \in \arcsMod} \height_a.
\end{align*}
\end{definition}

\vspace{2mm}
Since the traffic network under study is finite, and we consider only acyclic routes, we have $\height(G) < \infty$. Moreover, the case $\height(G) = 1$ corresponds to a parallel link network, for which the results of the following proposition have already been extensively analyzed in \cite{MaheshwariKulkarni2022DynamicTollingforInducingSociallyOptimalTrafficLoads}. We will henceforth assume that $\height(G) \geq 2$.

\begin{proposition} \label{Prop: Condensed DAG Properties, Height}
Given an Condensed DAG $\graphMod = (\nodesMod, \arcsMod)$ with the route set $\routes$:
\begin{enumerate}
    \item For any $a \in \arcsMod$, we have $\height_a = 1$ if and only if $j_a = d$. Similarly, if $\height_a =\height(G)$, then $i_a = o$.
    
    \item For any fixed $r \in \routes$, and any $a, a' \in r$ with $\height_{a, r} < \height_{a',r}$, we have $\height_a < \height_{a'}$ i.e., arcs along a route from the origin to the destination have strictly decreasing depth.
    
    \item Fix any $a \in \arcsMod$, and any $r \in \routes$ containing $a$ such that $\height_{a,r} = \height_a$. Then, for any $a' \in \routes$ following $a$ in $r$, we have $\height_{a',r} = \height_{a'}$.
    
    \item For each height $k \in [\height(\graphMod)] := \{1, \cdots, \height(\graphMod)\}$, there exists an arc $a \in \arcsMod$ such that $\height_a = k$.
\end{enumerate}
\end{proposition}

The proof of Proposition \ref{Prop: Condensed DAG Properties, Height} parallels that of Proposition \ref{Prop: Condensed DAG Properties, Depth}, and is omitted for brevity.

\subsection{Proofs of results in Section \ref{sec: New Equilibrium Characterization}}
\label{subsec: A2, App, Preliminaries}

\subsubsection{Proof of Lemma \ref{Prop: Strict Convexity of F}}
\label{subsubsec: A2, Proof, Strict Convexity of F}

Here, we establish Lemma \ref{Prop: Strict Convexity of F}, restated as follow: The map $F: \arcsLoadConstraintSet \ra \R$, as given below, is strictly convex.
\begin{align*}
    &F(\arcLoadMod) :=\sum_{[a] \in \arcsOrig} \int_0^{\arcLoadMod_{[a]}} \latency_{[a]}(u) \hspace{0.5mm} du 
    + \frac{1}{\beta} \sum_{i \ne d} \Bigg[ \sum_{a \in \arcsMod_i^+} \arcLoadMod_a \ln \arcLoadMod_a \\
    &\hspace{1.5cm} - \Bigg(\sum_{a \in \arcsMod_i^+} \arcLoadMod_a \Bigg) \ln \Bigg(\sum_{a \in \arcsMod_i^+} \arcLoadMod_a \Bigg) \Bigg].
\end{align*}

For convenience, we define $f_{[a]}: \arcsLoadConstraintSet \ra \R$, $\chi_i: \R^{|\arcsMod_i^+|} \ra \R$, $F: \arcsLoadConstraintSet \ra \R$ for each $[a] \in \arcsOrig$, $i \in \nodesMod \backslash \{d\}$ by:
\begin{align*}
    f_{[a]}(\arcLoadMod) &:= \int_0^{\arcLoadMod_{[a]}}  \latency_{[a]}(u) \hspace{0.5mm} du, \hspace{1cm} \forall \hspace{0.5mm} [a] \in \arcsOrig, \\
    \chi_i(\arcLoadMod_{\arcsMod_i^+}) &:= \sum_{a \in \arcsMod_i^+} \arcLoadMod_a \ln \arcLoadMod_a - \Bigg(\sum_{a \in \arcsMod_i^+} \arcLoadMod_a \Bigg) \ln \Bigg(\sum_{a \in \arcsMod_i^+} \arcLoadMod_a \Bigg), \\
    &\hspace{1cm} \forall \hspace{0.5mm} i \ne \nodesMod \backslash \{d\},
\end{align*}
where $\arcLoadMod_{\arcsMod_i^+} \in \R^{|\arcsMod_i^+|}$ denotes the components of $\arcLoadMod$ corresponding to arcs in $\arcsMod_i^+$. Then:
\begin{align*}
    F(\arcLoadMod) &= \sum_{[a] \in A_0} f_{[a]}(\arcLoadMod) + \frac{1}{\beta} \sum_{i \in \nodesMod \backslash \{d\}} \chi_i^\beta(\arcLoadMod).
\end{align*}

Also, for convenience, define:
\begin{align} \label{Eqn: Def, Ws}
    \arcsLoadConstraintSet_s &:= \Bigg\{ \arcLoadMod \in \R^{|\arcsMod|}: \sum_{a \in \arcsMod_i^+} \arcLoadMod_a = \sum_{a \in \arcsMod_i^-} \arcLoadMod_a, \hspace{0.5mm} \forall \hspace{0.5mm} i \ne o, d, \\ \nonumber
    &\hspace{1cm} \sum_{a \in \arcsMod_o^+} \arcLoadMod_a = 0. \Bigg\}.
\end{align}
Essentially, $\arcsLoadConstraintSet_s$ is the tangent space of the linear manifold with boundary $\arcsLoadConstraintSet$. Note that, using the notation described at the end of Section \ref{sec: Introduction and Related Works}, we can rewrite \eqref{Eqn: Def, Ws} as:
\begin{align} \nonumber
    \arcsLoadConstraintSet_s &= \big\{e_{\arcsMod_i^-} - e_{\arcsMod_i^+}: i \ne o, d \big\}^\perp \cap \big\{ e_{\arcsMod_o^+} \big\}^\perp.
\end{align}
We can now establish the strict convexity of $F$.



We first establish the convexity of $F$. It suffices to show that $f_{[a]}$ and $\chi_i$ are convex for each $[a] \in \arcsOrig$, $i \in \nodesMod \backslash \{d\}$. Note that each $f_{[a]}$ is convex since it is the composition of a convex function (\(g(w) = \sum_{a\in A_0}\int_{0}^{w_a}s_a(u) d u\)) with a linear function ($\arcLoadMod_{[a]} := \sum_{a' \in [a]} \arcLoadMod_{a'}$). We show below that $\chi_i$ is convex, for each $i \in \nodesMod \backslash \{d\}$.

Fix $i \in \nodesMod \backslash \{d\}$. For any $a, a' \in \arcsMod_i^+$ and each $\arcLoadMod \in \arcsLoadConstraintSet$:
\begin{align*}
    \frac{\partial^2 \chi_i}{\partial \arcLoadMod_a \partial \arcLoadMod_{a'}}(\arcLoadMod) &= \frac{1}{\arcLoadMod_a} \textbf{1}\{a' = a\} - \frac{1}{\sum_{\bar a \in \arcsMod_i^+} \arcLoadMod_{\bar a}}.
\end{align*}
Thus, for any $y \in \R^{|\arcsMod_i^+|}$:
\begin{align} \nonumber
    &\hspace{5mm} y^\top \nabla_\arcLoadMod^2 \chi_i(\arcLoadMod) y \\
    \nonumber
    &= \sum_{a, a' \in \arcsMod_i^+} y_a y_{a'} \frac{\partial^2 \chi_i}{\partial \arcLoadMod_a \partial \arcLoadMod_{a'}}(\arcLoadMod) \\ \nonumber
    &= \sum_{a \in \arcsMod_i^+} \frac{y_a^2}{\arcLoadMod_a} - \frac{1}{\sum_{\bar a \in \arcsMod_i^+} \arcLoadMod_{\bar a}} \cdot \sum_{a, a' \in \arcsMod_i^+} y_a y_{a'} \\ \nonumber
    &= \frac{1}{\sum_{\bar a \in \arcsMod_i^+} \arcLoadMod_{\bar a}} \left(\sum_{\bar a \in \arcsMod_i^+} \arcLoadMod_{\bar a} \cdot \sum_{a \in \arcsMod_i^+} \frac{y_a^2}{\arcLoadMod_a} - \left(\sum_{a' \in \arcsMod_i^+} y_{a'} \right)^2 \right) \\ \nonumber
    &= \frac{1}{\sum_{\bar a \in \arcsMod_i^+} \arcLoadMod_{\bar a}} \Bigg(\sum_{\bar a \in \arcsMod_i^+} \big(\sqrt{\arcLoadMod_{\bar a}} \big)^2 \cdot \sum_{a \in \arcsMod_i^+} \Bigg( \frac{y_a}{\sqrt{\arcLoadMod_a}} \Bigg)^2 \\ \nonumber
    &\hspace{1cm} - \left(\sum_{a' \in \arcsMod_i^+} \sqrt{\arcLoadMod_{a'}} \cdot \frac{y_{a'}}{\sqrt{\arcLoadMod_{a'}}} \right)^2 \Bigg) \\ \label{Eqn: Symmetric PSD test for chi i Hessian}
    &\geq 0,
\end{align}
where the final inequality follows from the Cauchy-Schwarz inequality. Cauchy-Schwarz also implies that equality holds in \eqref{Eqn: Symmetric PSD test for chi i Hessian} if and only if the vectors $(\sqrt{\arcLoadMod_a})_{a \in \arcsMod_i^+} \in \R^{|\arcsMod_i^+|}$ and $(y_a / \sqrt{\arcLoadMod_a})_{a \in \arcsMod_i^+} \in \R^{|\arcsMod_i^+|}$ are parallel, i.e., if $(y_a)_{a \in \arcsMod_i^+}$ and $(\arcLoadMod_a)_{a \in \arcsMod_i^+}$ are scalar multiples of each other. This shows that $\chi_i$ is convex, and $dim(N(\nabla_\arcLoadMod^2 \chi_i)) = 1$.


Second, suppose by contradiction that $F$ is not strictly convex on $\arcsLoadConstraintSet$. Then there exists some $\bar \arcLoadMod \in \arcsLoadConstraintSet$, $z \in \arcsLoadConstraintSet_s \backslash \{0\}$ such that:
\begin{align*}
    z^\top \nabla_\arcLoadMod^2 F(\bar \arcLoadMod) z = 0.
\end{align*}
Since $\nabla_\arcLoadMod^2F(\bar \arcLoadMod)$ is symmetric positive semidefinite, this is equivalent to stating that $z$ is in $N(\nabla_\arcLoadMod^2F(\bar \arcLoadMod))$, the null space of $\nabla_\arcLoadMod^2F(\bar \arcLoadMod)$. Let $\arcsMod_{z}$ denote the set of arc indices for which $z$ has a nonzero component, i.e.:
\begin{align*}
    \arcsMod_{z} := \{a' \in \arcsMod: z_{a'} \ne 0 \}.
\end{align*}
Since $z$ is not the zero vector, $\arcsMod_{z}$ is non-empty. Since there are a discrete and finite number of levels of $\graphMod$, there exists some $a \in \arcsMod_{z}$ such that $\ell_a \leq \ell_{a'}$ for all $a' \in \arcsMod_{y}$, i.e., $\ell_a = \min\{\ell_{a'}: a' \in \arcsMod_{y}\}$. Without loss of generality, we consider the case $z_a > 0$ (if not, then replace $z$ with $- z$, which would also be a nonzero vector in $N(\nabla_\arcLoadMod^2F(\bar \arcLoadMod))$).
We claim that $\arcLoadMod_a \ne 0$, and that for all $a' \in \arcsMod_{i_a}^+$:
\begin{align*}
    z_{a'} = z_a \cdot \frac{\arcLoadMod_{a'}}{\arcLoadMod_a} \geq 0.
\end{align*}
To see this, note that otherwise, the vectors $(z_a)_{a \in \arcsMod_i^+} \in \R^{|\arcsMod_i^+|}$ and $(\arcLoadMod_a)_{a \in \arcsMod_i^+}$ are not parallel, and so equality cannot be obtained in \eqref{Eqn: Symmetric PSD test for chi i Hessian}, i.e.,:
\begin{align*}
    z^\top\nabla_\arcLoadMod^2 \chi_i(\bar \arcLoadMod) z > 0,
\end{align*}
where, with a slight abuse of notation, we have defined $\chi_i(\arcLoadMod) = \chi_i(\arcsMod_i^+)$.
As a result:
\begin{align*}
    &\hspace{5mm} z^\top \nabla_\arcLoadMod^2 F(\bar \arcLoadMod) z \\
    &= \sum_{[a] \in \arcsMod} z^\top \nabla_\arcLoadMod^2 f_{[a]}(\bar \arcLoadMod) z + \frac{1}{\beta} \sum_{i' \ne d} z^\top \nabla_\arcLoadMod^2 \chi_{i'}(\bar \arcLoadMod) z \\
    &\geq \frac{1}{\beta} z^\top \nabla_\arcLoadMod^2 \chi_i(\bar \arcLoadMod)z \\
    &> 0,
\end{align*}
a contradiction. Thus, $z_a > 0$, and $z_{a'} \geq 0$ for each $a' \in \arcsMod_{i_a}^+$, so:
\begin{align*}
    \sum_{a' \in \arcsMod_{i_a}^+} z_{a'} > 0.
\end{align*}
If $\ell_a = 1$, i.e., $i_a = o$, we arrive at a contradiction, since the fact that $z \in \arcsLoadConstraintSet_s$ implies $\sum_{a' \in \arcsMod_{i_a}^+} z_{a'} = 0$. If $\ell_a > 1$, we also arrive at a contradiction, since the fact that $z \in \arcsLoadConstraintSet_s$ implies:
\begin{align*}
    \sum_{\hat a \in \arcsMod_{i_a}^-} z_{\hat a} = \sum_{a' \in \arcsMod_{i_a}^+} z_{a'} > 0,
\end{align*}
so there exists at least one $\ell_{\hat a} \in \arcsMod_{i_a}^-$ with $z_{\hat a} > 0$. Then, by definition of $a \in \arcsMod$, we have $\ell_a \leq \ell_{\hat a}$; this contradicts Proposition \ref{Prop: Condensed DAG Properties, Depth}, Part 2, which implies that since $\hat a \in \arcsMod_{i_a}^-$, there exists at least one arc containing $\hat a$ immediately before $a \in \arcsMod$, and thus $\ell_{\hat a} \leq \ell_a - 1$.
These contradictions complete the proof of the strict convexity of $F$ on $\arcsLoadConstraintSet$.

\subsubsection{Proof of Theo
rem \ref{Thm: CoDAG is unique minimizer of F}}
\label{subsubsec: A2, Proof, CoDAG is unique minimizer of F}

We present the proof of Theorem \ref{Thm: CoDAG is unique minimizer of F}, restated as follows: The Condensed DAG Equilibrium $\bar \arcLoadMod^\beta \in \arcsLoadConstraintSet$ exists, is unique, and is the unique optimal solution to the following convex optimization problem:
\begin{align*}
    \min_{w \in \arcsLoadConstraintSet} &\hspace{2mm} \sum_{[a] \in A_0} \int_0^{\arcLoadOrig_{[a]}} \latency_{[a]}(u) \hspace{0.5mm} dz \\
    &+ \frac{1}{\beta} \sum_{i \ne d} \Bigg[ \sum_{a \in A_i^+} \arcLoadMod_a \ln \arcLoadMod_a \\
    &\hspace{1cm} - \Bigg(\sum_{a \in A_i^+} \arcLoadMod_a \Bigg) \ln \Bigg(\sum_{a \in A_i^+} \arcLoadMod_a \Bigg) \Bigg].
\end{align*}

\begin{proof}(\textbf{Proof of Theorem \ref{Thm: CoDAG is unique minimizer of F}})
The following proof parallels that of Baillon, Cominetti \cite[Theorem 2]{BaillonCominetti2008MarkovianTrafficEquilibrium}. Recall that $N$ denotes the set of nodes of the corresponding DAG. The Lagrangian $\mathcal{L}: W \times \R^{|N| - 1} \in \R^{|A|} \ra \R$ corresponding to the above optimization problem is:
\begin{align*}
    &\hspace{5mm} \mathcal{L}(w, \mu, \lambda) \\
    &:= \sum_{[a] \in A_0} \int_0^{\arcLoadMod_{[a]}} \latency_{[a]}(u) \hspace{0.5mm} dz \\
    &\hspace{3mm} + \frac{1}{\beta} \sum_{i \ne d} \Bigg[ \sum_{a \in A_i^+} \arcLoadMod_a \ln \arcLoadMod_a - \Bigg(\sum_{a \in A_i^+} \arcLoadMod_a \Bigg) \ln \Bigg(\sum_{a \in A_i^+} \arcLoadMod_a \Bigg) \Bigg] \\
    &\hspace{3mm} + \sum_{i \ne d} \mu_i \Bigg( \nodeLoadIn_i + \sum_{a' \in A_i^-} \arcLoadMod_{a'} - \sum_{a' \in A_i^+} \arcLoadMod_{a'} \Bigg) + \sum_{a \in A} \lambda_a \arcLoadMod_a,
\end{align*}
with $\nodeLoadIn_i = \nodeLoadIn_o \cdot \textbf{1}\{i = o\}$, where $\textbf{1}\{\cdot\}$ is the indicator function that returns 1 if the input argument is true, and $0$ otherwise. At optimum $(w^\star, \mu^\star) \in \arcsLoadConstraintSet \times \R^{|N| - 1}$, the KKT conditions give, for each $a \in A$:
\begin{align*}
    0 &= \frac{\partial \mathcal{L}}{\partial \arcLoadMod_a} (w^\star, \mu^\star) \\
    &= \latency_{[a]}(\arcLoadMod_{[a]}^\star) + \frac{1}{\beta} \ln\left( \frac{\arcLoadMod_a^\star}{\sum_{a' \in A_{i_a}^+} \arcLoadMod_{a'}^\star} \right) + \mu_{j_a}^\star - \mu_{i_a}^\star + \lambda_a, \\
    0 &= \lambda_a \arcLoadMod_a, \hspace{5mm} \forall \hspace{0.5mm} a \in A.
\end{align*}

We claim that $(\hat \arcLoadMod, \hat \mu) \in \arcsLoadConstraintSet \times \R^{|N| - 1}$, as given by the Condensed DAG equilibrium definition: For each $a \in A$, $i \in N$:
\begin{align*}
    \hat \arcLoadMod_a &= \Bigg(\nodeLoadIn_{i_a} + \sum_{a' \in A_{i_a}^-} \hat \arcLoadMod_{a'} \Bigg) \cdot \frac{\exp(-\beta \costToGo_a(\hat \arcLoadMod))}{ \sum_{a' \in A_{i_a}^+} \exp(-\beta \costToGo_{a'}(\hat \arcLoadMod))}, \\
    &\hspace{1cm} \forall \hspace{0.5mm} a \in A, \\
    \hat \mu_i &= \nodeCostToGo_i(\costToGo(\hat \arcLoadMod)) = - \frac{1}{\beta} \ln\Bigg( \sum_{a' \in A_i^+} e^{-\beta \costToGo_{a'} (\hat \arcLoadMod)} \Bigg), \\
    &\hspace{1cm} \forall \hspace{0.5mm} i \in N, \\
    \hat \lambda_a &= 0, \hspace{5mm} \forall \hspace{0.5mm} a \in A,
\end{align*}
satisfies the KKT conditions stated above. Indeed, we have $\hat \arcLoadMod_a \geq 0$ for each $a \in A$, and:
\begin{align*}
    &\frac{\partial \mathcal{L}}{\partial \arcLoadMod_a}(\hat \arcLoadMod, \hat \mu, \hat 
    \lambda) \\
    = \hspace{0.5mm} &\latency_{[a]}(\hat \arcLoadMod_{[a]}) + \frac{1}{\beta} \ln\left( \frac{\hat \arcLoadMod_a}{\sum_{a' \in A_{i_a}^+} \hat \arcLoadMod_{a'}} \right) \\
    &\hspace{5mm} + \hat \mu_{j_a} - \hat \mu_{i_a} + \sum_{a \in A} \lambda_a \\
    = \hspace{0.5mm} &\latency_{[a]}(\hat \arcLoadMod_{[a]}) + \frac{1}{\beta} \ln\left( \frac{\exp(-\beta \costToGo_a(\arcLoadMod))}{ \sum_{a' \in A_{i_a}^+} \exp(-\beta \costToGo_{a'}(\arcLoadMod))} \right) \\
    &\hspace{5mm} + \nodeCostToGo_{j_a}(\costToGo) - \nodeCostToGo_{i_a}(\costToGo) \\
    = \hspace{0.5mm} &\latency_{[a]}(\hat \arcLoadMod_{[a]}) - \costToGo_a(\arcLoadMod) + \nodeCostToGo_{i_a}(\costToGo) + \nodeCostToGo_{j_a}(\costToGo) - \nodeCostToGo_{i_a}(\costToGo) \\
    = \hspace{0.5mm} &\latency_{[a]}(\hat \arcLoadMod_{[a]}) + \nodeCostToGo_{j_a}(\costToGo) - \costToGo_a(\arcLoadMod) \\
    = \hspace{0.5mm} &0,
\end{align*}
where the final equality follows from the definition of $(\costToGo_a)_{a \in A}$.
\end{proof}

\subsection{Proofs for Section \ref{sec: Methods}}
\label{subsec: A3, Methods}

\subsubsection{Proof of Lemma \ref{Lemma: Convergence, w, continuous}}
\label{subsubsec: A3, Proof, Convergence of w dynamics, continuous}

We present the proof of Lemma \ref{Lemma: Convergence, w, continuous}, stated formally as follows: Suppose $\arcLoadMod(0) \in \arcsLoadConstraintSet$, and:
\begin{align*}
    K_i > \frac{g_o}{C_w} \max\{K_{i_{\hat a}}: \hat a \in \arcsMod_i^-\}
\end{align*}
for each $i \in \nodesMod \backslash \{d\}$, with $C_w$ given by Lemma \ref{Lemma: Technical Conditions for Stochastic Approximation}. Then, the continuous-time dynamical system \eqref{Eqn: w flow, recursive, with h} for the traffic flow $\arcLoadMod(t)$ globally asymptotically converges to the corresponding Condensed DAG Equilibrium $\bar \arcLoadMod^\beta \in \arcsLoadConstraintSet$.

\begin{proof}(\textbf{Proof of Lemma \ref{Lemma: Convergence, w, continuous}})
We recursively write the continuous-time evolution of the arc flows $\arcLoadMod(\cdot)$ as follows, from \eqref{Eqn: General Network, xi flow, continuous} and \eqref{Eqn: General Network, w flow, continuous}. Recall that for any fixed $w \in \arcsLoadConstraintSet$, at each non-destination node $i \in \nodesMod \backslash \{d\}$, we have $\sum_{a' \in A_{i_a}^+} w_{a'} = \sum_{\hat a \in A_{i_a}^-} w_{\hat a}$. Thus, for each $a \not\in A_o^+$:
\begin{align} \nonumber
    &\dot \arcLoadMod_a(t) \\ \nonumber
    = \hspace{0.5mm} &\dot \probDist_a(t) \cdot \Bigg(g_{i_a} + \sum_{\hat a \in \arcsMod_{i_a}^-} \arcLoadMod_{\hat a}(t) \Bigg) + \probDist_a(t) \cdot \sum_{\hat a \in \arcsMod_{i_a}^{-}} \dot \arcLoadMod_{a}(t) \\ \nonumber
    = \hspace{0.5mm} &K_{i_a} \Bigg( - \probDist_a(t) 
    + \frac{\exp(-\beta \costToGo_a(\arcLoadMod(t)))}{\sum_{a' \in \arcsMod_{i_a}^+} \exp(-\beta \costToGo_{a'}(\arcLoadMod(t)))} \Bigg) \\ \nonumber
    &\hspace{1cm} \cdot \Bigg(g_{i_a} + \sum_{\hat a \in \arcsMod_{i_a}^-} \arcLoadMod_{\hat a}(t) \Bigg) \\ \nonumber
    &\hspace{5mm} + \probDist_a(t) \cdot \sum_{\hat a \in \arcsMod_{i_a}^{-}} \dot \arcLoadMod_{\hat a}(t) \\ \nonumber
    = \hspace{0.5mm} &- K_{i_a} \arcLoadMod_a(t) \\ \nonumber
    &\hspace{5mm} + K_{i_a} \cdot \Bigg(g_{i_a} + \sum_{\hat a \in \arcsMod_{i_a}^-} \arcLoadMod_{\hat a}(t) \Bigg) \\ \nonumber
    &\hspace{1cm} \cdot \frac{\exp(-\beta \costToGo_a(\arcLoadMod(t)))}{\sum_{a' \in \arcsMod_{i_a}^+} \exp(-\beta \costToGo_{a'}(\arcLoadMod(t)))} \\ \nonumber
    &\hspace{5mm} + \frac{\arcLoadMod_a(t)}{\sum_{a' \in \arcsMod_{i_a}^+} \arcLoadMod_{a'}(t)} \cdot \sum_{\hat a \in \arcsMod_{i_a}^-} \dot \arcLoadMod_{\hat a}(t) \\ \label{Eqn: w flow, recursive}
    = \hspace{0.5mm} &- K_{i_a} \Bigg( 1 - \frac{1}{K_{i_a}} \cdot \frac{ \sum_{\hat a \in \arcsMod_{i_a}^-} \dot \arcLoadMod_{\hat a}}{\sum_{a' \in \arcsMod_{i_a}^+} \arcLoadMod_{a'}} \Bigg) \arcLoadMod_a \\ \nonumber
    &\hspace{5mm} + K_{i_a} \cdot \Bigg(g_{i_a} + \sum_{\hat a \in \arcsMod_{i_a}^-} \arcLoadMod_{\hat a}(t) \Bigg) \\ \nonumber
    &\hspace{1cm} \cdot \frac{\exp(-\beta \costToGo_a(\arcLoadMod(t)))}{\sum_{a' \in \arcsMod_{i_a}^+} \exp(-\beta \costToGo_{a'}(\arcLoadMod(t)))}, 
\end{align}
for each $a \in \arcsMod$. More formally, we define each component $h: \arcsLoadConstraintSet \ra \R^{|\arcsMod|}$ recursively as follows. First, for each $a \in \arcsMod_o^+$, we set:
\begin{align*}
    h_a(\arcLoadMod) &:= K_o \left( -  \arcLoadMod_a + \nodeLoadIn_o \cdot \frac{\exp(-\beta \costToGo_a(\arcLoadMod))}{\sum_{a' \in \arcsMod_o^+} \exp(-\beta \costToGo_{a'}(\arcLoadMod))} \right).
\end{align*}
Suppose now that, for some arc $a \in A$, the component $h_a: \arcsLoadConstraintSet \ra \R$ of $h$ has been defined for each $\hat a \in \arcsMod_{i_a}^-$. Then, we set:
\begin{align*}
    h_a(\arcLoadMod) &:= - K_{i_a} \Bigg( 1 - \frac{1}{K_{i_a}} \cdot \frac{ \sum_{\hat a \in \arcsMod_{i_a}^-} h_{\hat a}(\arcLoadMod)}{\sum_{a' \in \arcsMod_{i_a}^+} \arcLoadMod_{a'}} \Bigg) \arcLoadMod_a \\ \nonumber
    &\hspace{7.5mm} + K_{i_a} \cdot \sum_{a' \in \arcsMod_{i_a}^-} \arcLoadMod_{a'} \cdot \frac{\exp(-\beta \costToGo_a(\arcLoadMod))}{\sum_{a' \in \arcsMod_o^+} \exp(-\beta \costToGo_{a'}(\arcLoadMod))}.
\end{align*}
By iterating through the above definition forward through the Condensed DAG $\graphMod$ from origin to destination, we can completely specify each $h_a$ in a well-posed manner (For a more rigorous characterization of this iterative procedure, see Appendix \ref{subsec: A1, Properties of Depth, Height}, Proposition \ref{Prop: Condensed DAG Properties, Depth}). We then define the $\arcLoadMod$-dynamics corresponding to the $\probDist$-dynamics \eqref{Eqn: General Network, xi flow, continuous} by:
\begin{align} \label{Eqn: w flow, recursive, with h}
    \dot \arcLoadMod &= h(\arcLoadMod).
\end{align}
Now, recall the objective $F: \arcsLoadConstraintSet \times \R^{|\arcsOrig|} \ra \R$ of the optimization problem that characterizes $\bar \arcLoadMod^\beta$, first stated in Theorem \ref{Thm: CoDAG is unique minimizer of F} as Equation \eqref{Eqn: Def, F}, reproduced below:
\begin{align} \nonumber
    &F(\arcLoadMod) \\ \nonumber
    := \hspace{0.5mm} &\sum_{[a] \in A_0} \int_0^{\arcLoadMod_{[a]}} \latency_{[a]}(\costToGo) \hspace{0.5mm} dz \\ \nonumber
    + &\frac{1}{\beta} \sum_{i \ne d} \Bigg[ \sum_{a \in A_i^+} \arcLoadMod_a \ln \arcLoadMod_a - \Bigg(\sum_{a \in A_i^+} \arcLoadMod_a \Bigg) \ln \Bigg(\sum_{a \in A_i^+} \arcLoadMod_a \Bigg) \Bigg].
\end{align}
Roughly speaking, our main approach is to show that \textit{$F$ is a Lyapunov function for the best-response dynamics in \eqref{Eqn: w flow, recursive, with h}}. Let $\arcsLoadConstraintSet_s$ denote the tangent space to $\arcsLoadConstraintSet$, and let $\Pi_{\arcsLoadConstraintSet_s}$ denote the orthogonal projection onto the linear subspace $\arcsLoadConstraintSet_s$. Under the continuous-time flow dynamics \eqref{Eqn: General Network, xi flow, continuous} and \eqref{Eqn: General Network, w flow, continuous}, if $\arcLoadMod \ne \bar \arcLoadMod^\beta$:
{\small
\begin{align} \nonumber
    &\hspace{5mm} \frac{d}{dt}(F \circ \arcLoadMod)(t) \\ \label{Eqn: dot w is in Ws, Expl required, 1}
    &= \dot \arcLoadMod(t)^\top \nabla_\arcLoadMod F(\arcLoadMod(t)) \\ \label{Eqn: dot w is in Ws, Expl required, 2}
    &= \dot \arcLoadMod(t)^\top \Pi_{\arcsLoadConstraintSet_s} \nabla_\arcLoadMod F(\arcLoadMod(t)) \\ \nonumber
    &= \dot \arcLoadMod(t)^\top  \Pi_{\arcsLoadConstraintSet_s} \big( \nabla_\arcLoadMod f(\arcLoadMod(t)) + \nabla \chi^\beta(\arcLoadMod(t)) \big) \\ \label{Eqn: Nabla f to nabla v, Expl required, 1}
    &= \dot \arcLoadMod(t)^\top \Pi_{\arcsLoadConstraintSet_s} \big( \big( \latency_{[a]}(\arcLoadMod_{[a]}(t)) \big)_{a \in \arcsMod} + \nabla \chi^\beta(\arcLoadMod(t)) \big) \\ \label{Eqn: Nabla f to nabla v, Expl required, 2}
    &= \dot \arcLoadMod(t)^\top \Pi_{\arcsLoadConstraintSet_s} \Bigg[ - \nabla \chi^\beta \Bigg( \Bigg( \Bigg( \nodeLoadIn_{i_a} + \sum_{a' \in \arcsMod_{i_a}^-} \arcLoadMod_{a'}(t) \Bigg) \\ \nonumber
    &\hspace{1cm} \cdot \frac{\exp(-\beta \costToGo_a(\arcLoadMod(t)))}{\sum_{\bar a \in \arcsMod_i^+} \exp(- \beta \costToGo_{\bar a} (\arcLoadMod(t)))} \Bigg)_{a \in \arcsMod} \Bigg) + \nabla \chi^\beta(\arcLoadMod(t)) \Bigg] \\ \label{Eqn: nabla v beta of w term, Expl required}
    &= \dot \arcLoadMod(t)^\top \Pi_{\arcsLoadConstraintSet_s} \Bigg[ - \nabla \chi^\beta \Bigg( \Bigg(  \Bigg( \nodeLoadIn_{i_a} + \sum_{a' \in \arcsMod_{i_a}^-} \arcLoadMod_{a'}(t) \Bigg) \\ \nonumber
    &\hspace{1.5cm} \cdot \frac{\exp(-\beta \costToGo_a(\arcLoadMod(t)))}{\sum_{\bar a \in \arcsMod_i^+} \exp(- \beta \costToGo_{\bar a} (\arcLoadMod(t)))} \Bigg)_{a \in \arcsMod} \Bigg) \\ \nonumber
    &\hspace{5mm} + \nabla \chi^\beta \left(\left( \left( 1 - \frac{\sum_{a' \in \arcsMod_{i_a}^-} h_{a'}(\arcLoadMod(t))}{K_{i_a} \cdot \sum_{\hat a \in \arcsMod_{i_a}^+} \arcLoadMod_{\hat a}(t)} \right) \cdot \arcLoadMod_a(t) \right)_{a \in \arcsMod} \right) \Bigg] \\ \label{Eqn: Final Expression, Expl required}
    &= \Bigg[ \Bigg( - K_{i_a} \Bigg( 1 - \frac{\sum_{a' \in \arcsMod_{i_a}^-} h_{a'}(\arcLoadMod(t))}{K_{i_a} \cdot \sum_{\hat a \in \arcsMod_{i_a}^+} \arcLoadMod_{\hat a}(t)} \Bigg) \arcLoadMod_a(t) \\ \nonumber
    &\hspace{1.5cm} + K_{i_a} \Bigg( \nodeLoadIn_{i_a} + \sum_{a' \in \arcsMod_{i_a}^+} \arcLoadMod_{a'}(t) \Bigg) \\ \nonumber
    &\hspace{2.5cm} \cdot \frac{\exp(-\beta \costToGo_a(\arcLoadMod(t)))}{\sum_{ \bar a \in \arcsMod_{i_a}^+} \exp(-\beta \costToGo_{\bar a} (\arcLoadMod(t)))} \Bigg)_{a \in \arcsMod} \Bigg]^\top \\ \nonumber
    &\hspace{5mm} \Bigg[ - \nabla \chi^\beta \Bigg( \Bigg(  \Bigg( \nodeLoadIn_{i_a} + \sum_{a' \in \arcsMod_{i_a}^-} \arcLoadMod_{a'}(t) \Bigg) \\ \nonumber
    &\hspace{3cm} \cdot \frac{\exp(-\beta \costToGo_a(\arcLoadMod(t)))}{\sum_{\bar a \in \arcsMod_i^+} \exp(- \beta \costToGo_{\bar a} (\arcLoadMod(t)))} \Bigg)_{a \in \arcsMod} \Bigg) \\ \nonumber
    &\hspace{7mm} + \nabla \chi^\beta \left(\left( \left( 1 - \frac{\sum_{a' \in \arcsMod_{i_a}^-} h_{a'}(\arcLoadMod(t))}{K_{i_a} \cdot \sum_{\hat a \in \arcsMod_{i_a}^+} \arcLoadMod_{\hat a}(t)} \right) \cdot \arcLoadMod_a(t) \right)_{a \in \arcsMod} \right) \Bigg] \\ \label{Eqn: Final Expression, Expl required, 2}
    &< 0.
\end{align}
}
We explain the equalities $\eqref{Eqn: dot w is in Ws, Expl required, 1} = \eqref{Eqn: dot w is in Ws, Expl required, 2}$, $\eqref{Eqn: Nabla f to nabla v, Expl required, 1} = \eqref{Eqn: Nabla f to nabla v, Expl required, 2}$, $\eqref{Eqn: Nabla f to nabla v, Expl required, 2} = \eqref{Eqn: nabla v beta of w term, Expl required}$, and $\eqref{Eqn: Final Expression, Expl required} < \eqref{Eqn: Final Expression, Expl required, 2}$ below. 

\paragraph{Verifying $\eqref{Eqn: dot w is in Ws, Expl required, 1} = \eqref{Eqn: dot w is in Ws, Expl required, 2}$}
From the equations leading up to \eqref{Eqn: w flow, recursive}, we have, for each $\arcLoadMod \in \arcsLoadConstraintSet$:
\begin{align*}
    \arcLoadMod_a &= \Bigg( g_{i_a} + \sum_{\hat a \in \arcsMod_{i_a}^-} \arcLoadMod_{a'} \Bigg) \cdot \probDist_a,
\end{align*}
and so:
\begin{align*}
    &\hspace{5mm} \dot \arcLoadMod_a(t) \\
    &= \Bigg( \nodeLoadIn_{i_a} + \sum_{\hat a \in \arcsMod_{i_a}^-} \arcLoadMod_{\hat a}(t) \Bigg) \cdot \dot \probDist_a + \sum_{\hat a \in \arcsMod_{i_a}^-} \dot \arcLoadMod_{a'} \cdot \probDist_a \\
    &= \Bigg( \nodeLoadIn_{i_a} + \sum_{\hat a \in \arcsMod_{i_a}^-} \arcLoadMod_{\hat a}(t) \Bigg) \\
    &\hspace{1.5cm} \cdot K_{i_a} \left( - \probDist_a + \frac{\exp(-\beta \costToGo_a(\arcLoadMod(t)))}{\sum_{a' \in \arcsMod_{i_a}^+} \exp(-\beta \costToGo_{a'}(\arcLoadMod(t)))} \right) \\
    &\hspace{1cm} + \sum_{\hat a \in \arcsMod_{i_a}^-} \dot \arcLoadMod_{a'} \cdot \probDist_a
\end{align*}
Fix any node $i \in \nodesMod$ in the Condensed DAG, and consider the sum of the above equation over the arc subset $\arcsMod_i^+$:
\begin{align*}
    \sum_{a' \in \arcsMod_i^+} \dot \arcLoadMod_{a'}(t)
    &= \sum_{\hat a \in \arcsMod_i^-} \dot \arcLoadMod_{\hat a}(t).
\end{align*}
Since $\arcLoadMod(0) \in \arcsLoadConstraintSet$ by assumption, we have the initial condition $(\sum_{\hat a \in \arcsMod_i^+} \arcLoadMod_{\hat a} - \sum_{a' \in \arcsMod_{i_a}^-} \arcLoadMod_{a'} - \nodeLoadIn_i)(0) = 0$ for the above linear time-invariant differential equation. We thus conclude that, for each $t \geq 0$:
\begin{align*}
    \sum_{\hat a \in \arcsMod_i^+} \arcLoadMod_{\hat a}(t) - \sum_{a' \in \arcsMod_{i_a}^-} \arcLoadMod_{a'}(t) - \nodeLoadIn_i  = 0.
\end{align*}
Since this holds for any arbitrary node $i \in \nodesMod$, we have $\arcLoadMod(t) \in \arcsLoadConstraintSet$ for all $t \geq 0$.

\paragraph{Verifying $\eqref{Eqn: Nabla f to nabla v, Expl required, 1} = \eqref{Eqn: Nabla f to nabla v, Expl required, 2}$} 

We will show that:
\begin{align} \label{Eqn: Nabla f to nabla v, to show, 2}
    &\Pi_{\arcsLoadConstraintSet_s} \Bigg[ \big( \latency_{[a]}(\arcLoadMod_{[a]}(t)) \big)_{a \in \arcsMod} + \nabla \chi^\beta \Bigg( \Bigg(  \Bigg( \nodeLoadIn_{i_a} + \sum_{a' \in \arcsMod_{i_a}^-} \arcLoadMod_{a'}(t) \Bigg) \\ \nonumber
    &\hspace{1cm} \cdot \frac{\exp(-\beta \costToGo_a(\arcLoadMod(t)))}{\sum_{\bar a \in \arcsMod_i^+} \exp(- \beta \costToGo_{\bar a} (\arcLoadMod(t)))} \Bigg)_{a \in \arcsMod} \Bigg) \Bigg] = 0,
\end{align}
which would a fortiori establish the desired claim $\eqref{Eqn: Nabla f to nabla v, Expl required, 1} = \eqref{Eqn: Nabla f to nabla v, Expl required, 2}$. To do so, first note that, for each $i \ne d$, $a \in \arcsMod_i^+$: 
\begin{align} \label{Eqn: Nabla v beta wrt w}
    \frac{\partial \chi^\beta}{\partial \arcLoadMod_a} (\arcLoadMod) &= \frac{1}{\beta} \cdot \Bigg[ \ln \arcLoadMod_a + 1 - \ln\left( \sum_{a \in \arcsMod_i^+} \arcLoadMod_a \right) - 1 \Bigg] \\ \nonumber
    &= \frac{1}{\beta} \ln\left( \frac{\arcLoadMod_a}{\sum_{a \in \arcsMod_i^+} \arcLoadMod_a} \right).
\end{align}
Thus, we have:
\begin{align*}
    &\frac{\partial \chi^\beta}{\partial \arcLoadMod_a}  \Bigg( \Bigg(  \Bigg( \nodeLoadIn_{i_a} + \sum_{a' \in \arcsMod_{i_a}^-} \arcLoadMod_{a'} \Bigg) \\
    &\hspace{1.5cm} \cdot \frac{\exp(-\beta \costToGo_a(\arcLoadMod))}{\sum_{\bar a \in \arcsMod_{i_a}^+} \exp(- \beta \costToGo_{\bar a} (\arcLoadMod))} \Bigg)_{a \in \arcsMod} \Bigg) \\
    = \hspace{0.5mm} &\frac{1}{\beta} \ln \left( \frac{\exp(-\beta \costToGo_a(\arcLoadMod))}{\sum_{\bar a \in \arcsMod_i^+} \exp(- \beta \costToGo_{\bar a} (\arcLoadMod))} \right) \\
    = \hspace{0.5mm} & - \costToGo_a(\arcLoadMod) - \frac{1}{\beta} \ln \left( \sum_{\bar a \in \arcsMod_{i_a}^+} \exp(- \beta \costToGo_{\bar a} (\arcLoadMod)) \right) \\
    = \hspace{0.5mm} & - \costToGo_a(\arcLoadMod) + \nodeCostToGo_{i_a}(\arcLoadMod).
\end{align*}
Concatenating these partial derivatives to form the gradient, we can now verify \eqref{Eqn: Nabla f to nabla v, to show, 2} by observing that:
\begin{align*}
    &\Pi_{\arcsLoadConstraintSet_s} \Bigg[ \big( \latency_{[a]}(\arcLoadMod_{[a]}) \big)_{a \in \arcsMod} \\
    &\hspace{1cm} + \nabla \chi^\beta \Bigg( \Bigg(  \Bigg( \nodeLoadIn_{i_a} + \sum_{a' \in \arcsMod_{i_a}^-} \arcLoadMod_{a'} \Bigg) \\
    &\hspace{2cm} \cdot \frac{\exp(-\beta \costToGo_{\hat a}(\arcLoadMod))}{\sum_{\bar a \in \arcsMod_i^+} \exp(- \beta \costToGo_{\bar a} (\arcLoadMod))} \Bigg)_{\hat a \in \arcsMod} \Bigg)_{a \in \arcsMod} \Bigg] \\
    = \hspace{0.5mm} &\Pi_{\arcsLoadConstraintSet_s} \big( \latency_{[a]}(\arcLoadMod_{[a]}) - \costToGo_a(\arcLoadMod) + \nodeCostToGo_{i_a}(\arcLoadMod) \big)_{a \in \arcsMod} \\
    = \hspace{0.5mm} &\Pi_{\arcsLoadConstraintSet_s} \big( \nodeCostToGo_{i_a}(\arcLoadMod) - \nodeCostToGo_{j_a}(\arcLoadMod) \big)_{a \in \arcsMod} \\
    = \hspace{0.5mm} &\Pi_{\arcsLoadConstraintSet_s} \Bigg[ \sum_{a \in \arcsMod} \nodeCostToGo_{i_a}(\arcLoadMod) e_a - \sum_{a \in \arcsMod} \nodeCostToGo_{j_a}(\arcLoadMod) e_a \Bigg] \\
    = \hspace{0.5mm} &\Pi_{\arcsLoadConstraintSet_s} \Bigg[ - \sum_{\hat a \in \arcsMod_d^-} \nodeCostToGo_{j_{\hat a}}(\arcLoadMod) e_{\hat a} + \sum_{a' \in \arcsMod_o^+} \nodeCostToGo_{i_{a'}}(\arcLoadMod) e_{a'} \\
    &\hspace{1cm} + \sum_{i \ne \{o, d\}} \Bigg(\sum_{a' \in \arcsMod_i^+} \nodeCostToGo_i(\arcLoadMod) e_{a'} - \sum_{\hat a \in \arcsMod_i^-} \nodeCostToGo_i(\arcLoadMod) e_{\hat a} \Bigg) \Bigg] \\
    = \hspace{0.5mm} &\Pi_{\arcsLoadConstraintSet_s} \Bigg[ 0 + \nodeCostToGo_o(\arcLoadMod) e_{\arcsMod_o^+} + \sum_{i \ne \{o, d\}} \nodeCostToGo_i(\arcLoadMod) \big(e_{\arcsMod_i^-} - e_{\arcsMod_i^+} \big) \Bigg] \\
    = \hspace{0.5mm} &0,
\end{align*}
where the last equality follows by definition of $\Pi_{\arcsLoadConstraintSet_s}$.

\paragraph{Verifying $\eqref{Eqn: Nabla f to nabla v, Expl required, 2} = \eqref{Eqn: nabla v beta of w term, Expl required}$} 

We will show that:
\begin{align*}
    \nabla \chi^\beta(\arcLoadMod) = \nabla \chi^\beta \left(\left( \left( 1 - \frac{\sum_{a' \in \arcsMod_{i_a}^-} \dot \arcLoadMod_{a'}}{K_{i_a} \cdot \sum_{\hat a \in \arcsMod_{i_a}^+} \arcLoadMod_{\hat a}} \right) \cdot \arcLoadMod_a \right)_{a \in \arcsMod} \right), 
\end{align*}
which is equivalent to showing that $\eqref{Eqn: Nabla f to nabla v, Expl required, 2} = \eqref{Eqn: nabla v beta of w term, Expl required}$. From \eqref{Eqn: Nabla v beta wrt w}, we have for each $a \in \arcsMod$:
\begin{align*}
    &\frac{\partial \chi^\beta}{\partial \arcLoadMod_a} \left(\left( \left( 1 - \frac{\sum_{a' \in \arcsMod_{i_a}^-} h_{a'}(\arcLoadMod)}{K_{i_a} \cdot \sum_{\hat a \in \arcsMod_{i_a}^+} \arcLoadMod_{\hat a}} \right) \cdot \arcLoadMod_a \right)_{a \in \arcsMod} \right) \\
    = \hspace{0.5mm} &\frac{1}{\beta} \ln \left(\frac{\left( 1 - \frac{\sum_{a' \in \arcsMod_{i_a}^-} h_{a'}(\arcLoadMod)}{K_{i_a} \cdot \sum_{\hat a \in \arcsMod_{i_a}^+} \arcLoadMod_{\hat a}} \right) \arcLoadMod_a}{\sum_{\bar a \in \arcsMod_{i_a}^+} \left( 1 - \frac{\sum_{a' \in \arcsMod_{i_{\bar a}}^-} h_{a'}(\arcLoadMod)}{K_{i_a} \cdot \sum_{\hat a \in \arcsMod_{i_{\bar a}}^+} \arcLoadMod_{\hat a}} \right) \arcLoadMod_{\bar a}} \right) \\
    = \hspace{0.5mm} &\frac{1}{\beta} \ln \left(\frac{\left( 1 - \frac{\sum_{a' \in \arcsMod_{i_a}^-} h_{a'}(\arcLoadMod)}{K_{i_a} \cdot \sum_{\hat a \in \arcsMod_{i_a}^+} \arcLoadMod_{\hat a}} \right) \arcLoadMod_a}{ \left( 1 - \frac{\sum_{a' \in \arcsMod_{i_a}^-} h_{a'}(\arcLoadMod)}{K_{i_a} \cdot \sum_{\hat a \in \arcsMod_{i_a}^+} \arcLoadMod_{\hat a}} \right) \cdot \sum_{\bar a \in \arcsMod_{i_a}^+} \arcLoadMod_{\bar a}} \right) \\
    = \hspace{0.5mm} &\frac{1}{\beta} \ln \left(\frac{\arcLoadMod_a}{ \sum_{\bar a \in \arcsMod_{i_a}^+} \arcLoadMod_{\bar a}} \right) \\
    = \hspace{0.5mm} & \frac{\partial \chi^\beta}{\partial \arcLoadMod_a}(\arcLoadMod).
\end{align*}
The second equality above follows because, for each $\bar a \in \arcsMod_{i_a}^+$, we have $i_{\bar a} = i_a$. This verifies that $\eqref{Eqn: Nabla f to nabla v, Expl required, 2} = \eqref{Eqn: nabla v beta of w term, Expl required}$.

\paragraph{Verifying $\eqref{Eqn: Final Expression, Expl required} < \eqref{Eqn: Final Expression, Expl required, 2}, \forall \hspace{0.5mm} \arcLoadMod \ne \bar \arcLoadMod^\beta$}

Suppose $\frac{d}{dt}(F \circ \arcLoadMod) = 0$ at some $\tilde \arcLoadMod \in \arcsLoadConstraintSet$. From \eqref{Eqn: Final Expression, Expl required}, and by the definition of the convex function $\chi$:
{\small
\begin{align*}
    &\hspace{5mm} 0 \\
    &= \frac{d}{dt}(F \circ w) \\
    &= \sum_{i \in \nodesMod \backslash \{d\}} \Bigg( \Bigg[ - \Bigg(1 - \frac{\sum_{\hat a \in \arcsMod_{i_a}^-} h_{\hat a}(\arcLoadMod)}{K_{i_a} \cdot \sum_{a' \in \arcsMod_{i_a}^+} \tilde \arcLoadMod_{a'}} \Bigg) \tilde \arcLoadMod_a \\
    &\hspace{3mm} + \Bigg( g_{i_a} + \sum_{a' \in \arcsMod_{i_a}^+} \tilde \arcLoadMod_{a'} \Bigg) \frac{\exp(-\beta \cdot \costToGo_a(\tilde \arcLoadMod))}{\sum_{a' \in \arcsMod_{i_a}^+} \exp(-\beta \cdot \costToGo_{a'}(\tilde \arcLoadMod)) } \Bigg]_{a \in A} \Bigg)^\top \\
    &\hspace{1cm} \cdot \frac{1}{\beta} \Bigg( \nabla \chi_i^\beta\Bigg( \Bigg[ \Bigg(1 - \frac{\sum_{\hat a \in \arcsMod_{i_a}^-} h_{\hat a}(\tilde \arcLoadMod)}{K_{i_a} \cdot \sum_{a' \in \arcsMod_{i_a}^+} \tilde \arcLoadMod_{a'}} \Bigg) \tilde \arcLoadMod_a \Bigg]_{a \in \arcsMod} \Bigg) \\
    &\hspace{1cm} - \nabla \chi_i^\beta \Bigg( \Bigg[ \Bigg( g_{i_a} + \sum_{a' \in \arcsMod_{i_a}^+} \tilde \arcLoadMod_{a'} \Bigg) \\
    &\hspace{3cm} \cdot \frac{\exp(-\beta \cdot \costToGo_a(\tilde \arcLoadMod))}{\sum_{a' \in \arcsMod_{i_a}^+} \exp(-\beta \cdot \costToGo_{a'}(\tilde \arcLoadMod)) } \Bigg]_{a \in \arcsMod} \Bigg),
\end{align*}
}
where, for each $i \in \nodesMod \backslash \{d\}$, the convex map $\chi_i^\beta: \R^{|\arcsMod_i^+|} \ra \R$ is defined by:
\begin{align*}
    &\chi_i^\beta(\{\arcLoadMod_a: a \in \arcsMod_i^+ \}) \\
    = \hspace{0.5mm} &\sum_{a \in \arcsMod_i^+} \arcLoadMod_a \ln \arcLoadMod_a - \Bigg( \sum_{a \in \arcsMod_i^+} \arcLoadMod_a \Bigg) \ln \Bigg( \sum_{a \in \arcsMod_i^+} \arcLoadMod_a \Bigg).
\end{align*}
The convexity of each $\chi_i^\beta$ implies that each of the above summands must be non-positive; since they sum to 0, each summand must be 0. (By assumption, $K_{i_a} > (g_o/C_w) \max\{K_{i_{\hat a}}: \hat a \in A_{i_a}^- \}$, so the input arguments to th $\nabla \chi_i(\cdot)$ maps are all strictly positive.)

Now, for each $i \in \nodesMod \backslash \{d\}$ and each $\arcLoadMod \in \R^{\arcsMod_i^+}$, we have $N(\nabla^2 \chi_i^\beta(w)) = \text{span}\{w\}$. In words, $\chi_i^\beta$ increases linearly only along rays emanating from the origin. In the context of the above summands, this implies that, for each $i \in \nodesMod \backslash \{d\}$, there exists constants $Q_i \in \R$ such that, for each $a \in \arcsMod_i^+$:
\begin{align*}
    &Q_i \Bigg(1 - \frac{\sum_{\hat a \in \arcsMod_{i_a}^-} h_{\hat a}(\arcLoadMod)}{K_{i_a} \cdot \sum_{a' \in \arcsMod_{i_a}^+} \arcLoadMod_{a'}} \Bigg) \arcLoadMod_a \\
    = \hspace{0.5mm} &\Bigg( g_{i_a} + \sum_{a' \in \arcsMod_{i_a}^+} \arcLoadMod_{a'} \Bigg) \cdot \frac{\exp(-\beta \cdot \costToGo_a(\arcLoadMod))}{\sum_{a' \in \arcsMod_{i_a}^+} \exp(-\beta \cdot \costToGo_{a'}(\arcLoadMod)) }.
\end{align*}
By definition of $h: \arcsLoadConstraintSet \ra \R^{|\arcsMod|}$, for each $a \in \arcsMod_o^+$:
\begin{align*}
    h_a(\arcLoadMod) &= - \tilde \arcLoadMod_a + \nodeLoadIn_o \cdot \frac{\exp(-\beta \costToGo_a(\arcLoadMod))}{\sum_{a' \in \arcsMod_o^+} \exp(-\beta \costToGo_{a'}(\arcLoadMod))} \\
    &= (Q_o - 1) \arcLoadMod_a
\end{align*}
and for each $a \in \arcsMod_i^+$ with $i \ne o$:
\begin{align*}
    h_a(\arcLoadMod) &:= - \Bigg( 1 - \frac{ \sum_{\hat a \in \arcsMod_{i_a}^-} h_{\hat a}(\arcLoadMod)}{\sum_{a' \in \arcsMod_{i_a}^+} \arcLoadMod_{a'}} \Bigg) \arcLoadMod_a \\ \nonumber
    &\hspace{1cm} + \sum_{a' \in \arcsMod_{i_a}^-} \arcLoadMod_{a'} \cdot \frac{\exp(-\beta \costToGo_a(\arcLoadMod))}{\sum_{a' \in \arcsMod_o^+} \exp(-\beta \costToGo_{a'}(\arcLoadMod))} \\
    &= (Q_o - 1) \Bigg( 1 - \frac{ \sum_{\hat a \in \arcsMod_{i_a}^-} h_{\hat a}(\arcLoadMod)}{\sum_{a' \in \arcsMod_{i_a}^+} \arcLoadMod_{a'}} \Bigg) \arcLoadMod_a.
\end{align*}
By the flow continuity equations:
\begin{align*}
    0 = \sum_{a' \in \arcsMod_o^+} h_{a'}(\arcLoadMod) = (Q_o - 1) \cdot \sum_{a' \in \arcsMod_i^+} \arcLoadMod_{a'} = (Q_o - 1) g_o,
\end{align*}
so $Q_o = 1$, and thus $h_a(\arcLoadMod) = 0$ for each $a \in \arcsMod_o^+$. Now, suppose there exists some depth $\depth \in [\depth(G) - 1]$ such that $\depth_a(\arcLoadMod) = 0$ for each $a \in \arcsMod$ such that $\depth_a \leq \depth$. Then, for each $a \in \arcsMod$ such that $\depth_a = \depth+1$, the flow continuity equations give:
\begin{align*}
    0 &= \sum_{a' \in \arcsMod_i^+} h_{a'}(\arcLoadMod) - \sum_{\hat a \in \arcsMod_i^-} h_{\hat a}(\arcLoadMod) \\
    &= \sum_{a' \in \arcsMod_i^+} h_{a'}(\arcLoadMod) \\
    &= (Q_{i_a} - 1) \cdot \sum_{a' \in \arcsMod_{i_a}^+} \arcLoadMod_{a'}.
\end{align*}
Thus, $Q_{i_a} = 1$, so $h_a(\arcLoadMod) = 0$. This completes the recursion step, and shows that $h(\arcLoadMod) = 0$, i.e., $\arcLoadMod = \bar \arcLoadMod^\beta$.

In summary, we established that the map $F$ strictly decreases along any trajectory that starts in $\arcsLoadConstraintSet \backslash \{\bar \arcLoadMod^\beta\}$ and follows the best-response dynamics \eqref{Eqn: w flow, recursive, with h}. The convergence of the dynamics \eqref{Eqn: w flow, recursive, with h} to the Condensed DAG equilibrium \eqref{Def: CoDAG Equilibrium} now follows by invoking either Sandholm, Corollary 7.B.6 \cite{Sandholm2010PopulationGamesAndEvolutionaryDynamics}, or Sastry, Proposition 5.22 and Theorem 5.23 (LaSalle's Principle and its corollaries) \cite{Sastry1999NonlinearSystems}. 
\end{proof}


\subsubsection{Proof of Lemma \ref{Lemma: Technical Conditions for Stochastic Approximation}} 

To prove Lemma \ref{Lemma: Technical Conditions for Stochastic Approximation}, we require the following results. We first establish bounds on the trajectory of discrete-time and continuous-time traffic flow dynamics. 

\begin{lemma} \label{Lemma: Bounding Dynamics} ~\
\begin{enumerate}
    \item Consider the discrete-time dynamics:
    \begin{align*}
        Y[n+1] &= (1 - \delta[n]) Y[n] + \delta[n] F[n],
    \end{align*}
    where, for each $n \geq 0$, we have $\delta[n] \in (0, 1)$ and $Y[0], F[n] \in [a, b]$ for some $a, b \in \R$ satisfying $a < b$. Then $Y[n] \in [a, b]$ for each $n \geq 0$.
    
    \item Consider the continuous-time dynamics:
    \begin{align*}
        \dot y(t) &= K(- y(t) + f(t)),
    \end{align*}
    where $K > 0$ and, for each $t \geq 0$, we have $y(0), f(t) \in [a, b]$ for some $a, b \in \R$ satisfying $a < b$. Then $y(t) \in [a, b]$ for each $t \geq 0$.
\end{enumerate}
\end{lemma}

\begin{proof} ~\
\begin{enumerate}
    \item Suppose there exists some $N \geq 0$ such that $Y[n] \in [a, b]$ for each $n \leq N$. (Since $Y[0] \in [a, b]$ by assumption, this is certainly true for $n = 0$). Then:
    \begin{align*}
        Y[n+1] &= (1 - \delta[n]) Y[n] + \delta[n] F[n] \\
        &\in [(1 - \delta[n]) \cdot a + \delta[n] \cdot a, (1 - \delta[n]) \cdot b \\
        &\hspace{1cm} + \delta[n] \cdot b] \\
        &= [a, b].
    \end{align*}
    This completes the induction step, and thus the proof of this part of the lemma.
    
    \item We compute:
    \begin{align*}
        \frac{d}{dt}(e^{Kt} y(t)) &= e^{Kt} \big( \dot y(t) + a y(t) \big) = a e^{Kt} f(t).
    \end{align*}
    Integrating from time $0$ to time $t$, we have, for each $t \geq 0$:
    \begin{align*}
        &e^{Kt} y(t) - y(0) = \int_0^t a e^{K\tau} f(\tau) \hspace{0.5mm} d\tau.
    \end{align*}
    Rearranging terms, we obtain, for each $t \geq 0$:
    \begin{align*}
        y(t) &= e^{-Kt} y(0) + e^{-Kt} \int_0^t a e^{a \tau} f(\tau) \hspace{0.5mm} d\tau \\
        &\in \big[e^{-Kt} a + (1 - e^{-Kt}) a, e^{-Kt} b + (1 - e^{-Kt}) b \big] \\
        &= [a, b].
    \end{align*}

\end{enumerate}
\end{proof}

Before proceeding, we rewrite the discrete $\xi$-dynamics \eqref{Eqn: General Network, xi flow, discrete} as a Markov process with a martingale difference term:
\begin{align*}
    &\probDist_a[n+1] \\
    = \hspace{0.5mm} &\probDist_a[n] \\
    &\hspace{5mm} + \mu \Bigg(K_{i_a} \Bigg( -\probDist_a[n] + \frac{\exp(-\beta \big[ \costToGo_a(\arcLoadModDiscrete[n]) \big])}{\sum\limits_{ a' \in \arcsMod_{i_a}^+} \exp(-\beta \big[ \costToGo_{a'}(\arcLoadModDiscrete[n]) \big])} \Bigg) \\
    &\hspace{1cm} + M_a[n+1] \Bigg),
\end{align*}
with:
\begin{align} \nonumber
    &M_a[n+1] \\ \nonumber
    := \hspace{0.5mm} &\left( \frac{1}{\mu} \eta_{i_a}[n+1] - 1 \right) \\ \nonumber
    &\hspace{1mm} \cdot K_{i_a} \Bigg(- \probDist_a[n] + \frac{\exp(-\beta \big[ \costToGo_a(\arcLoadModDiscrete[n]) \big])}{\sum_{ a' \in \arcsMod_{i_a}^+} \exp(-\beta \big[ \costToGo_{a'}(\arcLoadModDiscrete[n]) \big])} \Bigg).
\end{align}
The following lemma bounds the magnitude of the discrete-time flow $\arcLoadModDiscrete[n] \in \R^{|\arcLoadMod|}$ and the martingale difference terms $M[n] \in \R^{|\arcLoadMod|}$.

\begin{lemma} \label{Lemma: Boundedness of Maps}
The discrete-time dynamics \eqref{Eqn: General Network, xi flow, discrete} and \eqref{Eqn: General Network, w flow, discrete} satisfy:
\begin{enumerate}
    \item For each $a \in \arcsMod$: $\{M_a[n+1]: n \geq 0\}$ is a martingale difference sequence with respect to the filtration $\mathcal{F}_n := \sigma\big( \cup_{a \in \arcsMod} (\arcLoadModDiscrete_a[1], \probDist[1], \cdots, \arcLoadModDiscrete_a[n], \probDist[n]) \big)$.
    
    \item There exist $C_w, C_m > 0$ such that, for each $a \in \arcsMod$ and each $n \geq 0$, we have:
    \begin{align*}
        \arcLoadModDiscrete_a[n] &\in [C_w, g_o], \\
        |M_a[n]| &\leq C_m.
    \end{align*}
\end{enumerate}
The continuous-time dynamics \eqref{Eqn: General Network, xi flow, continuous} and \eqref{Eqn: General Network, w flow, continuous} satisfy:
\begin{enumerate}[resume]
    \item For each $a \in \arcsMod$, $n \geq 0$:
    \begin{align*}
        \arcLoadMod_a(t) &\in [C_w, g_o].
    \end{align*}
\end{enumerate}
\end{lemma}

\begin{proof} ~\
\begin{enumerate}
    \item We have:
    \begin{align*}
        &\E[M_a[n+1] | \mathcal{F}_n] \\
        = \hspace{0.5mm} &\left( \frac{1}{\mu} \E[\eta_{i_a}[n+1]] - 1 \right)\\
        &\hspace{1mm} \cdot K_{i_a} \left(-\probDist_a[n] + \frac{\exp(-\beta \big[ \costToGo_a(\arcLoadModDiscrete[n]) \big])}{\sum_{ a' \in \arcsMod_{i_a}^+} \exp(-\beta \big[ \costToGo_{a'}(\arcLoadModDiscrete[n]) \big])} \right) \\
        = \hspace{0.5mm} &0.
    \end{align*}
    
    \item We separate the proof of this part of the lemma into the following steps. 
    \begin{itemize}
        \item First, we show that for each $a \in \arcsMod$, $n \geq 0$, we have $\probDist_a[n] \in (0, 1]$.
        
        $\hspace{5mm}$ Fix $a \in \arcsMod$ arbitrarily. Then $\probDist_a[0] \in (0, 1]$ by assumption, and for each $n \geq 0$:
        \begin{align*}
            \frac{\exp(-\beta \big[ \costToGo_a(\arcLoadModDiscrete[n]) \big])}{\sum_{ a' \in \arcsMod_{i_a}^+} \exp(-\beta \big[ \costToGo_{a'}(\arcLoadModDiscrete[n]) \big])} \in (0, 1],
        \end{align*}
        since the exponential function takes values in $(0, \infty)$. Thus, by Lemma \ref{Lemma: Bounding Dynamics}, we have $\probDist_a[n] \in (0, 1]$ for each $n \geq 0$.
        
        \item Second, we show that for each $a \in \arcsMod$, $n \geq 0$, we have $\arcLoadModDiscrete_a[n] \in (0, g_o]$.
        
        $\hspace{5mm}$ Note that \eqref{Eqn: General Network, w flow, discrete}, together with the assumption that $\arcLoadModDiscrete[0] \in \arcsLoadConstraintSet$, implies that $\arcLoadModDiscrete[n] \in \arcsLoadConstraintSet$ for each $n \geq 0$. Now, fix $a \in \arcsMod$, $n \geq 0$ arbitrarily. Let $\routes(a) \subseteq \routes$ denote the set of all routes passing through $a$, and for each $r \in \routes(a)$, let $a_{r, k}$ denote the $k$-th arc in $r$. Then, by the conservation of flow encoded in \routes:
        \begin{align*}
            \arcLoadModDiscrete_a[n] &= g_o \cdot \sum_{r \in \routes(a)} \prod_{k=1}^{|r|} \probDist_{a_{r,k}} \\
            &\leq g_o \cdot \sum_{r \in \routes} \prod_{k=1}^{|r|} \probDist_{a_{r,k}} \\
            &= g_o.
        \end{align*}
        Similarly, since $\probDist_a[n] \in (0, 1]$ for each $a \in \arcsMod$, $n \geq 0$, we have:
        \begin{align*}
            \arcLoadModDiscrete_a[n] &= g_o \cdot \sum_{r \in \routes(a)} \prod_{k=1}^{|r|} \probDist_{a_{r,k}} > 0.
        \end{align*}
        
        \item Third, we show that there exists $C_z > 0$ such that $|\costToGo_a(\arcLoadModDiscrete[n])| \leq C_z$ for each $a \in \arcsMod$, $n \geq 0$. Fix $a \in \arcsMod_d^- = \{a \in \arcsMod: m_a = 1\}$ arbitrarily. Then, from \eqref{Eqn: CostToGo}:
        \begin{align*}
            &\costToGo_a(\arcLoadMod) = \latency_{[a]}(\arcLoadMod_{[a]}) \in [0, \latency_{[a]}(g_o) ], \\
            \Ra \hspace{0.5mm} &|\costToGo_a(\arcLoadMod)| \leq \latency_{[a]}(g_o)  := C_{z,1}.
        \end{align*}
        Now, suppose that at some height $k \in [\height(\graphMod) - 1]$, there exists some $C_{z,k} > 0$ such that, for each $n \geq 0$, and each $a \in \arcsMod$ satisfying $\height_a \leq k$ and each $n \geq 0$, we have $|\costToGo_a(\arcLoadMod)| \leq C_{z,k}$. Then, for each $n \geq 0$, and each $a \in \arcsMod$ satisfying $\height_a = k+1$ (at least one such $a \in \arcsMod$ must exist, by Proposition \ref{Prop: Condensed DAG Properties, Height}, Part 4):
        \begin{align*}
            \costToGo_a(\arcLoadMod) &= \latency_{[a]}(\arcLoadMod_{[a]}) - \frac{1}{\beta} \ln\Bigg( \sum_{a' \in \arcsMod_{j_a}^+} e^{-\beta \cdot \costToGo_{a'}(\arcLoadMod)} \Bigg) \\
            &\leq \latency_{[a]}(g_o)  - \frac{1}{\beta} \ln\left( |\arcsMod_{j_a}^+| e^{-\beta \cdot C_z} \right) \\
            &= \latency_{[a]}(g_o)  + C_z,
        \end{align*}
        and:
        \begin{align*}
            \costToGo_a(\arcLoadMod) &= \latency_{[a]}(\arcLoadMod_{[a]}) - \frac{1}{\beta} \ln\Bigg( \sum_{a' \in \arcsMod_{j_a}^+} e^{-\beta \cdot \costToGo_{a'}(\arcLoadMod)} \Bigg) \\
            &\geq 0 + 0 - \frac{1}{\beta} \ln\left( |\arcsMod_{j_a}^+| e^{\beta \cdot C_z} \right) \\
            &= - \frac{1}{\beta}\ln|\arcsMod| - C_z ,
        \end{align*}
        from which we conclude that:
        \begin{align*}
            |\costToGo_a(\arcLoadMod)| &\leq \max\left\{\latency_{[a]}(g_o)  + C_z, \frac{1}{\beta}\ln|\arcsMod| + C_z \right\} \\
            &:= C_{z,k+1},
        \end{align*}
        with $C_{z+1} \geq C_z$. This completes the induction step, and the proof is completed by taking $C_z := C_{z, \height(\graphMod)}$.
        
        \item Fourth, we show that there exists some $C_\probDist > 0$ such that $\probDist_a[n] \geq C_\probDist$ for each $a \in \arcsMod$, $n \geq 0$. 
        
        $\hspace{5mm}$ Define:
        \begin{align*}
            C_\probDist &:= \min\left\{ \min_{a' \in \arcsMod} \probDist_{a'}[0], \frac{1}{|\arcsMod|} e^{-2 \beta C_z} \right\} \in (0, 1).
        \end{align*}
        By definition of $C_\probDist$, we have $\probDist_a[0] \geq C_\probDist$. Moreover, for each $n \geq 0$, we have:
        \begin{align*}
            &\frac{\exp(-\beta \big[ \costToGo_a(\arcLoadModDiscrete[n]) \big])}{\sum_{ a' \in \arcsMod_{i_a}^+} \exp(-\beta \big[ \costToGo_{a'}(\arcLoadModDiscrete[n]) \big])} \\
            \geq \hspace{0.5mm} &\frac{e^{-\beta C_z}}{|\arcsMod_{i_a}^+| \cdot e^{\beta C_z}} \\
            \geq \hspace{0.5mm} & \frac{1}{|\arcsMod|} e^{-2\beta C_z} \\
            \geq \hspace{0.5mm} &C_\probDist.
        \end{align*}
        Thus, by Lemma \ref{Lemma: Bounding Dynamics},  $\probDist_a[n] \geq C_\probDist$ for each $n \geq 0$.
        
        \item Fifth, we show that there exists $C_w > 0$ such that, for each $a \in \arcsMod$, $n \geq 0$, we have $\arcLoadModDiscrete_a[n] \geq C_w$.
        
        $\hspace{5mm}$ Fix $a \in \arcsMod$, $n \geq 0$. Let $\routes(a)$ denote the set of all routes in the Condensed DAG containing $a$, and let $r \in \routes(a)$ be arbitrarily given. By unwinding the recursive definition of $\arcLoadModDiscrete_a[n]$ from the flow dispersion probability values $\{\probDist_a[n]: a \in \arcsMod, n \geq 0\}$, we have:
        \begin{align*}
            \arcLoadModDiscrete_a[n] &= g_o \cdot \sum_{\substack{r' \in \routes \\ a \in r'}} \prod_{a' \in r'} \probDist_{a'}[n] \\
            &\geq g_o \cdot \prod_{a' \in r} \probDist_{a'}[n] \\
            &\geq g_o \cdot (C_\probDist)^{|r|} \\
            &\geq g_o \cdot (C_\probDist)^{\depth(\graphMod)} \\
            &:= C_w.
        \end{align*}
        
        \item Sixth, we show that there exists $C_m > 0$ such that, for each $a \in \arcsMod$, $n \geq 0$, we have $M_a[n] \geq C_m$.
        
        $\hspace{5mm}$ Define, for convenience, $C_\mu := \max\{\overline \mu - \mu, \mu - \underline \mu\}$. Since $\eta_{i_a}[n] \in [\underline \mu, \overline \mu]$, we have from \eqref{Eqn: Ma, Discrete-Time Dynamics} that for each $a \in \arcsMod$, $n \geq 0$:
        { \small
        \begin{align} \nonumber
                &M_a[n+1] \\ \nonumber
                := \hspace{0.5mm} &\left( \frac{1}{\mu} \eta_{i_a}[n+1] - 1 \right) \\ \nonumber
                &\hspace{1mm} \cdot K_{i_a} \Bigg(- \probDist_a[n] + \frac{\exp(-\beta \big[ \costToGo_a(\arcLoadModDiscrete[n]) \big])}{\sum_{ a' \in \arcsMod_{i_a}^+} \exp(-\beta \big[ \costToGo_{a'}(\arcLoadModDiscrete[n]) \big])} \Bigg).
        \end{align}
        }
        Thus, by the triangle inequality:
        \begin{align*}
            |M_a[n+1]| &\leq \frac{1}{\mu} C_\mu K_{i_a} \cdot (1+1) \\
            &\leq \frac{2}{\mu} C_\mu \cdot \max_{i \in \nodesMod \backslash \{d\}} K_i \\
            &:= C_m.
        \end{align*}
    \end{itemize}
    
    \item We separate the proof of this part of the lemma into the following steps.
    
    \begin{itemize}
        \item First, we show that for each $a \in \arcsMod$, $t \geq 0$, we have $\probDist_a(t) \in (0, 1]$.
        
        $\hspace{5mm}$ Fix $a \in \arcsMod$. By assumption, $\probDist_a(0) \in (0, 1]$, and at each $t \geq 0$:
        \begin{align*}
            \frac{\exp(-\beta \costToGo_a(\arcLoadMod))}{\sum_{a' \in \arcsMod_{i_a}^+} \exp(-\beta \costToGo_{a'}(\arcLoadMod))} \in (0, 1].
        \end{align*}
        Thus, by Lemma \ref{Lemma: Bounding Dynamics}, we conclude that $\probDist_a(t) \in (0, 1]$ for each $t \geq 0$.
        
        \item Second, we show that $\arcLoadMod_a(t) \in [0, g_o]$ for each $t \geq 0$.
        
        $\hspace{5mm}$ The proof here is nearly identical to the proof that $\arcLoadModDiscrete_a[n] \in (0, g_o)$ in the second bullet point of the second part of this lemma, and is omitted for brevity.
        
        \item Third, we show that $|\costToGo_a(\arcLoadMod_a(t))| \leq C_z$ for each $t \geq 0$.
        
        $\hspace{5mm}$ The proof here is nearly identical to the proof that $|\costToGo_a(\arcLoadModDiscrete_a[n])| \leq C_z$ in the fourth bullet point of the second part of this lemma, and is omitted for brevity.
        
        \item Fourth, we show that there exists some $C_\probDist > 0$ such that $\probDist_a(t) \geq C_\probDist$ for each $a \in \arcsMod$, $t \geq 0$.  
        
        $\hspace{5mm}$ Define:
        \begin{align*}
            C_\probDist &:= \min\left\{ \min\{\probDist_{a'}(0): a' \in \arcsMod\}, \frac{1}{|\arcsMod|} e^{-2 \beta C_z} \right\} \\
            &\in (0, 1).
        \end{align*}
        By definition of $C_\probDist$, we have $\probDist_a(0) \geq C_\probDist$. Moreover, for each $n \geq 0$, we have:
        \begin{align*}
            &\frac{\exp(-\beta \big[ \costToGo_a(\arcLoadModDiscrete[n]) \big])}{\sum_{ a' \in \arcsMod_{i_a}^+} \exp(-\beta \big[ \costToGo_{a'}(\arcLoadModDiscrete[n]) \big])} \\
            \geq \hspace{0.5mm} &\frac{e^{-\beta C_z}}{|\arcsMod_{i_a}^+| \cdot e^{\beta C_z}} \\
            \geq \hspace{0.5mm} & \frac{1}{|\arcsMod|} e^{-2\beta C_z} \\
            \geq \hspace{0.5mm} &C_\probDist.
        \end{align*}
        Thus, by Lemma \ref{Lemma: Bounding Dynamics}, we have $\probDist_a(t) \geq C_\probDist$ for each $t \geq 0$.
 
        \item Fifth, we show that there exists $C_w > 0$ such that, for each $a \in \arcsMod$, $t \geq 0$, we have $\arcLoadMod_a(t) \geq C_w$.
        
        $\hspace{5mm}$ The proof here is nearly identical to the proof that $\arcLoadModDiscrete_a[n] \geq C_w$ in the fourth bullet point of the second part of this lemma, and is omitted for brevity.
        
    \end{itemize}
\end{enumerate}
\end{proof}

\begin{remark}
Crucially, the constants introduced and used in the above proof, i.e., $C_w, C_m, C_\xi$ (and naturally, $g_o$), do not depend on the node-dependent update rates $K_i$. This is a critical observation, since each $K_i$ must be chosen to be large enough such that the term:
\begin{align*}
    1 - \frac{\sum_{a' \in \arcsMod_{i_a}^-} h_{a'}(\arcLoadMod(t))}{K_{i_a} \cdot \sum_{\hat a \in \arcsMod_{i_a}^+} \arcLoadMod_{\hat a}(t)}
\end{align*}
which appears in \eqref{Eqn: Final Expression, Expl required}, is always strictly positive, i.e., that:
\begin{align} \label{Eqn: Lower Bound on Kia}
    K_{i_a} > \frac{\sum_{\hat a \in \arcsMod_{i_a}^+} \arcLoadMod_{\hat a}(t)}{\sum_{a' \in \arcsMod_{i_a}^-} h_{a'}(\arcLoadMod(t))}
\end{align}
for all $t \geq 0$, regardless of the initial value of $w(0) \in \arcsLoadConstraintSet$. The numerator in the right-hand-side expression of \eqref{Eqn: Lower Bound on Kia} can be straightforwardly (if loosely) upper bounded by $|\arcsMod| g_o$. However, the denominator in the right-hand-side expression of \eqref{Eqn: Lower Bound on Kia} must be lower-bounded recursively in increasing order of depth, which requires $K_{i_a}$ to depend on $\{K_i: i \in \nodesMod, \ell_i < \ell_{i_a} \}$, as well as on the constants $C_w, C_m, C_\xi$, and $g_o$. Thus, the fact that $C_w, C_m, C_\xi$, and $g_o$ are established independently of the values of $K_i$ allows circular reasoning to be avoided.
\end{remark}

The lemma below establishes the final part of Lemma \ref{Lemma: Technical Conditions for Stochastic Approximation}. Below, we restrict the domains of the maps $\bar \probDist^\beta$ and $\costToGo_a$ to reflect the bounds of the traffic flow trajectory $\arcLoadMod$ established in the above lemma, i.e., $\bar \probDist^\beta, \costToGo_a: \arcsLoadConstraintSet' \ra \R$, with the flow restricted to:
\begin{align*}
    \arcsLoadConstraintSet' := \arcsLoadConstraintSet \cap [C_w, g_o]^{|\arcsMod|}
\end{align*}
and the toll restricted to $[0, C_p]^{|\arcsOrig|}$.

\begin{lemma} \label{Lemma: Lipschitz Continuity of Maps}
The continuous-time dynamics \eqref{Eqn: w flow, recursive, with h} satisfies:
\begin{enumerate}
    \item For each $a \in \arcsMod$, the restriction of the latency-to-go map $\costToGo_a: \arcsLoadConstraintSet \ra \R^{|\arcsOrig|} \ra \R$ to $\arcsLoadConstraintSet'$ is Lipschitz continuous.
    
    \item The map from the probability transitions $\probDist \in \prod_{i \in I \backslash \{d\}} \Delta(\arcsMod_i^+)$ and the traffic flows $\arcLoadMod \in \arcsLoadConstraintSet$ is Lipschitz continuous.
    
    \item For each $a \in \arcsMod$, the restriction of the continuous dynamics transition map $\rho_a: \R^{|\arcsMod|} \times \R^{|\arcsOrig|} \ra \R^{|\arcsMod|}$, defined recursively by:
    \begin{align*}
        \rho_a(\probDist) &:= K_{i_a} \left( - \probDist_a
        + \frac{\exp(-\beta \costToGo_a(\arcLoadMod))  }{\sum_{ a' \in \arcsMod_{i_a}^+} \exp(-\beta \costToGo_{a'}(\arcLoadMod))} \right)
    \end{align*}
    to $\arcsLoadConstraintSet'$ is Lipschitz continuous.
\end{enumerate}
\end{lemma}

\begin{proof} ~\
\begin{enumerate} 
    \item We shall establish the Lipschitz continuity of $\costToGo_a$, for each $a \in \arcsMod$, by providing uniform bounds on its partial derivatives across all values of its arguments $\arcLoadMod \in \arcsLoadConstraintSet' $.
    
    $\hspace{5mm}$ The proof follows by induction on the height index $k \in [\height(\graphMod)]$. For each $a \in \arcsMod$, let $\tilde \costToGo_a: \R^{|\arcsMod|} \ra \R$ be the continuous extension of $\costToGo_a: \arcsLoadConstraintSet \ra \R$ to the Euclidean space $\R^{|\arcsMod|}$ containing $\arcsLoadConstraintSet$. By definition of Lipschitz continuity, if $\tilde \costToGo_a$ is Lipschitz for some $a \in \arcsMod$, then so is $\costToGo_a$. For each $a \in \arcsMod_d^- = \{a \in \arcsMod: \height_a = 1\}$ and any $\arcLoadMod \in \R^{|\arcsMod|}$:
    \begin{align*}
        \tilde \costToGo_a(\arcLoadMod) = \latency_{[a]}(\arcLoadMod_{[a]}).
    \end{align*}
    Thus, for any $\hat a \in \arcsMod$, and any $\arcLoadMod \in \R^{|\arcsMod|}$, $p \in \R^{|\arcsOrig|}$:
    \begin{align*}
        \frac{\partial \tilde \costToGo_a}{\partial \arcLoadMod_{\hat a}}(\arcLoadMod) &= \frac{d\latency_{[a]}}{d\arcLoadMod}(\arcLoadMod_{[a]}) \cdot \textbf{1}\{\hat a \in [a]\} \in [0, C_{ds}].
    \end{align*}
    We set $C_{z,1} := C_{ds}$.
    
    $\hspace{5mm}$ Now, suppose that there exists some depth $k \in [\height(\graphMod) - 1]$ and some constant $C_{z,k} > 0$ such that, for any $a \in \arcsMod$ satisfying $\height_a \leq k$, and any $\arcLoadMod \in \arcsLoadConstraintSet$, $n \geq 0$, the map $\tilde \costToGo_a: \R^{|\arcsMod|} \ra \R$ is continuously differentiable, with:
    \begin{align*}
        \left|\frac{\partial \tilde \costToGo_a}{\partial \arcLoadMod_{\hat a}}(\arcLoadMod) \right| &\leq C_{z,k}.
    \end{align*}
    Continuing with the induction step, fix $a \in \arcsMod$ such that $\height_a = k+1$ (there exists at least one such link, by Proposition \ref{Prop: Condensed DAG Properties, Depth}, Part 4). From Proposition \ref{Prop: Condensed DAG Properties, Depth}, Part 2, we have $\height_{a'} \leq k$ for each $a' \in \arcsMod_{i_a}^+$. Thus, the induction hypothesis implies that, for any $\hat a \in \arcsMod$:
    \begin{align*}
        \tilde \costToGo_a(\arcLoadMod) &= \latency_{[a]}(\arcLoadMod_{[a]}) - \frac{1}{\beta} \sum_{a' \in \arcsMod_{j_a}^+} e^{-\beta \costToGo_{a'}(\arcLoadMod)}.
    \end{align*}
    Computing partial derivatives with respect to each component of $\arcLoadMod$, we obtain:
    \begin{align*}
        \frac{\partial \tilde \costToGo_a}{\partial \arcLoadMod_{\hat a}}(\arcLoadMod) &= \frac{d\latency_{[a]}}{d\arcLoadMod}(\arcLoadMod_{[a]}) \cdot \textbf{1}\{\hat a \in [a]\} \\
        &\hspace{1cm} + \sum_{a' \in \arcsMod_{j_a}^+} e^{-\beta \tilde \costToGo_{a'}(\arcLoadMod)} \cdot \frac{\partial \tilde \costToGo_{a'}}{\partial \arcLoadMod_{\hat a}}(\arcLoadMod), \\
        \Ra \hspace{0.5mm} \left| \frac{\partial \tilde \costToGo_a}{\partial \arcLoadMod_{\hat a}}(\arcLoadMod) \right| &\leq C_{ds} + |\arcsMod| \cdot C_{z,k}.
    \end{align*}
    We can complete the induction step by taking $C_{z,k+1} := C_{ds} + |\arcsMod| \cdot C_{z,k}$.
    
    $\hspace{5mm}$ This establishes that, for each $a \in \arcsMod$, the map $\costToGo_a$ is continuously differentiable, with partial derivatives uniformly bounded by a uniform constant, $C_z := C_{z,\height(\graphMod)}$. This establishes the Lipschitz continuity of the map $\costToGo_a$ for each $a \in \arcsMod$, and thus proves this part of the proposition.
    
    \item Recall that the map from traffic distributions probabilities ($\probDist$) to traffic flows ($\arcLoadMod$) is given as follows, for each $a \in \arcsMod$. Recall that $\routes(a)$ denotes the set of all routes in the Condensed DAG that contain the arc $a$:
    \begin{align*}
        \arcLoadMod_a = \left(g_{i_a} + \sum_{\hat a \in \arcsMod_i^-} \arcLoadMod_a \right) \cdot \probDist_a = g_o \cdot \sum_{r \in \routes(a)} \prod_{k=1}^{|r|} \probDist_{a_{r,k}},
    \end{align*}
    where $a_{r,k}$ denotes the $k$-th arc along a given route $r \in \routes$, for each $k \in |r|$. Thus, the map from $\probDist$ to $\arcLoadMod$ is continuously differentiable. Moreover, the domain of this map is compact; indeed, for each $a \in \arcsMod$, we have $\probDist_a \in [0, 1]$, and for each non-destination node $i \ne d$, we have $\sum_{a \in \arcsMod_i^+} \probDist_a = 1$. Therefore, the map $\probDist \mapsto \arcLoadMod$ has continuously differentiable derivatives with magnitude bounded above by some constant uniform in the compact set of realizable probability distributions $\probDist$. This is equivalent to stating that the map $\probDist \mapsto \arcLoadMod$ is Lipschitz continuous.
    
    \item Above, we have established that the maps $\costToGo_a$ and $\probDist \mapsto \arcLoadMod$ are Lipschitz continuous. Since the addition and composition of Lipschitz maps is Lipschitz, it suffices to verify that the map $\hat \rho: \R^{|\arcsMod|} \ra \R^{|\arcsMod|}$, defined element-wise by:
    \begin{align*}
        \hat \rho_a(z) := \frac{e^{-\beta \costToGo_a}}{\sum_{a' \in \arcsMod_{i_a}^+} e^{-\beta \costToGo_{a'}}}, \hspace{1cm} \forall \hspace{0.5mm} a \in \arcsMod
    \end{align*}
    is Lipschitz continuous. We do so below by computing, and establishing a uniform bound for, its partial derivatives. For each $\hat a \in \arcsMod$:
    \begin{align*}
        \frac{\partial \hat \rho_a}{\partial \costToGo_{\bar a}} &= \frac{1}{(\sum_{a' \in \arcsMod_{i_a}^+} e^{-\beta \costToGo_{a'}})^2} \\
        &\hspace{5mm} \cdot \Bigg( \sum_{a' \in \arcsMod_{i_a}^+} e^{-\beta \costToGo_{a'}} \cdot (-\beta) e^{-\beta \costToGo_a} \cdot \frac{\partial \costToGo_a}{\partial \costToGo_{\bar a}} \\
        &\hspace{1.5cm} - e^{-\beta \costToGo_a} \cdot \sum_{a' \in \arcsMod_{i_a}^+} (-\beta) e^{-\beta \costToGo_{a'}} \frac{\partial \costToGo_{a'}}{\partial \costToGo_{\bar a}} \Bigg) \\
        &= - \frac{e^{-\beta \costToGo_a}}{\sum_{a' \in \arcsMod_{i_a}^+} e^{-\beta \costToGo_{a'}}} \cdot \beta \cdot \frac{\partial \costToGo_a}{\partial \costToGo_{\hat a}} \\
        &\hspace{7mm}+ \frac{\beta e^{-\beta \costToGo_a}}{(\sum_{a' \in \arcsMod_{i_a}^+} e^{-\beta \costToGo_{a'}})^2} \cdot \sum_{a' \in \arcsMod_{i_a}^+} e^{-\beta \costToGo_{a'}} \frac{\partial \costToGo_{a'}}{\partial \costToGo_{\bar a}}.
    \end{align*}
    Observe that:
    \begin{align*}
        \sum_{a' \in \arcsMod_{i_a}^+} e^{-\beta \costToGo_{a'}} \frac{\partial \costToGo_{a'}}{\partial \costToGo_{\bar a}} &= \sum_{a' \in \arcsMod_{i_a}^+} e^{-\beta \costToGo_{a'}} \cdot \textbf{1}\{a' = \hat a\} \\
        &\leq \max_{a' \in \arcsMod_{i_a}^+} e^{-\beta \costToGo_{a'}}.
    \end{align*}
    This, together with triangle inequality, then gives:
    \begin{align*}
        \left| \frac{\partial \hat \rho_a}{\partial \costToGo_{\bar a}} \right| &= \beta + \beta = 2\beta.
    \end{align*}
    This concludes the proof for this part of the proposition.
    
\end{enumerate}
\end{proof}

We present the proof of Theorem \ref{Thm: Convergence, w, discrete}, restated as follows: For any $\delta > 0$:
\begin{align*}
    \limsup_{n \ra \infty} \E\big[ \Vert \probDist[n] - \bar \probDist^\beta \Vert_2^2 \big] &\leq O(\mu), \\
    \limsup_{n \ra \infty} \Prob\big( \Vert \probDist[n] - \bar \probDist^\beta \Vert_2^2 \geq \delta \big) &\leq O\left( \frac{\mu}{\delta} \right).
\end{align*}

\subsubsection{Proof of Theorem \ref{Thm: Convergence, w, discrete}}
\label{subsubsec: A3, Proof, Convergence of w dynamics, discrete}

Here, we conclude the proof of Theorem \ref{Thm: Convergence, w, discrete}.

\begin{proof}(\textbf{Proof of Theorem \ref{Thm: Convergence, w, discrete}})
Lemma \ref{Lemma: Boundedness of Maps} asserts that $M[n]$ is bounded (uniformly in $n \geq 0$), while Lemma \ref{Lemma: Lipschitz Continuity of Maps} establishes that $\rho: \R^{|A|} \ra \R$ is Lipschitz continuous. The proof of Theorem \ref{Thm: Convergence, w, discrete} now follows by applying the stochastic approximation results in Borkar \cite{Borkar2008StochasticApproximation}, Chapters 2 and 9.
\end{proof}

\end{document}